\documentclass[structabstract]{aa}  
\usepackage{float}
\usepackage{natbib}
\usepackage{graphicx}
\usepackage{txfonts}
\begin{document}
   \title{Enhanced H$_2$O formation through dust grain chemistry in X-ray exposed environments}

   \author{R. Meijerink\inst{1,2} \and S. Cazaux\inst{2} \and M. Spaans\inst{2} }

   \institute{Leiden Observatory, Leiden University, P.O. Box 9513,
     NL-2300 RA Leiden, Netherlands \and
     Kapteyn Astronomical Institute, PO Box 800, 9700 AV Groningen, The Netherlands\\
     \email{meijerink@astro.rug.nl}
    }
   \date{Received ??; accepted ??}

   \abstract{The Ultra-luminous infrared galaxy Mrk 231, showing signs
     of both black hole accretion and star formation, exhibits very
     strong water rotational lines between $\lambda = 200 -
     670$~$\mu$m, comparable to the strength of the CO rotational
     lines. High redshift quasars also show similar CO and H$_2$O line
     properties, while starburst galaxies, such as M82, lack these
     very strong H$_2$O lines in the same wavelength range, but do
     show strong CO lines.}{We explore the possibility of enhancing
     the gas phase H$_2$O abundance in X-ray exposed environments,
     using bare interstellar carbonaceous dust grains as a
     catalyst. Cloud-cloud collisions cause C and J shocks, and strip
     the grains of their ice layers. The internal UV field created by
     X-rays from the accreting black hole does not allow to reform the
     ice.}{We determine formation rates of both OH and H$_2$O on dust
     grains, having temperature $T_{\rm dust}=10-60$~K, using both
     Monte Carlo as well as rate equation method simulations. The
     acquired formation rates are added to our X-ray chemistry code,
     that allows us to calculate the thermal and chemical structure of
     the interstellar medium near an active galactic nucleus.}{We
     derive analytic expressions for the formation of OH and H$_2$O on
     bare dust grains as a catalyst. Oxygen atoms arriving on the dust
     are released into the gas phase under the form of OH and
     H$_2$O. The efficiencies of this conversion due to the chemistry
     occurring on dust are of order 30 percent for oxygen converted
     into OH and 60 percent for oxygen converted into H$_2$O between
     $T_{\rm dust}=15-40$~K. At higher temperatures, the efficiencies
     rapidly decline. When the gas is mostly atomic, molecule
     formation on dust is dominant over the gas-phase route, which is
     then quenched by the low H$_2$ abundance.  Here, it is possible
     to enhance the warm ($T> 200$~K) water abundance by an order of
     magnitude in X-ray exposed environments. This helps to explain
     the observed bright water lines in nearby and high-redshift
     ULIRGs and Quasars.}{}


   \maketitle
%

\section{Introduction}

Warm molecular gas ($T > 100$~K) cools mainly through CO, H$_2$O, and
H$_2$ ro-vibrational transitions \citep{Neufeld1993,Neufeld1995}. The
ground state lines of CO, up to $J=6-5$, are extensively studied from
the ground in a large number of galaxies \citep[see,
  e.g.,][]{Papadopoulos2010} up to a redshift $z\sim
0.04$. Unfortunately, the atmosphere makes it impossible to study CO
transitions with $J > 8$ and H$_2$O rotational lines. It is these
lines, however, that are tracing the warm molecular gas. Moreover,
these lines are sensitive to extreme UV irradiation from
star-formation and X-rays due to supermassive black-hole
accretion. Some high-$J$ CO lines have been observed in bright
galactic objects such as the Orion Bar and Sagittarius A
\citep[cf.][]{Hollenbach1999}, and the Class 0 object L1448-mm
\citep{Nisini1999}. It is only recently that significant progress is
being made in the study of the warm molecular interstellar medium
(ISM) in nearby galaxies. Both high-$J$ as well as rotational water
lines have been observed abundantly after the launch of the Herschel
Space observatory, that is operating between wavelengths
$\lambda=55-672$~$\mu$m. The successful HerCULES open time key program
has observed 29 close (U)LIRGs with both SPIRE and PACS instruments,
revealing CO lines up CO $J=14-13$ (while current follow-up PACS
observations might detect even higher transition) and H$_2$O lines in
Mrk 231 that are of comparable strengths
\citep{VanderWerf2010,Gonzalez2010}. These CO and H$_2$O lines are at
least in part produced in regions of the galaxy exposed to X-rays,
so-called XDRs \citep{Maloney1996,Meijerink2005}, but emission
produced through shock excitation may also contribute as observed for
OH in Mrk 231 \citep{Fischer2010}. Moreover, {\it Herschel}-PACS also
detected a lot of high-$J$ CO lines as well as H$_2$O and OH in
embedded objects, such as DK Chamaeleontis \citep{VanKempen2010} and
in disk atmospheres, e.g., HD 100546 \citep{Sturm2010}, where X-rays
also affect the chemical-thermal structure. Even though this paper
focuses on an application to ULIRGs, it will have applications to
other objects as well.

Understanding the excitation and chemistry of these important
molecules is therefore key in studying the dominant physical processes
in the ISM and star-formation. CO is a linear molecule, and the
radiation transfer models allow to calculate far-infrared to
millimeter spectra relatively easy for a given physical
structure. Models predicting CO emission, where both the
thermal-chemical structure and radiation transfer are calculated
simultaneously have more problems to reach agreement. The case for
H$_2$O is even worse, due to its complex nature: high critical
densities $n_{\rm crit} > 10^8 - 10^{10}$~cm$^{-3}$ in combination
with high opacities, infrared-pumping and maser actively, makes an
excitation calculation extremely challenging. Also, the excitation
state in which water will enter the gas-phase after desorption or
evaporation from a dust grain is very uncertain. A first guess in this
would be to divide the excess energy over the levels using
equipartion, so 1/3 translational, 1/3 rotational and 1/3 vibrational
excitation, with a Boltzmann distribution for the sub-levels. The
formation of molecular hydrogen, H$_2$, in the ISM has been a
long-standing problem. At extremely low metallicities, when no dust is
present, H$_2$ either forms through the H$^-$ route \citep[$\rm H^- +
  H \rightarrow H_2 + e^{-1}$;][]{Glover2003,Launey1991}, or through a
three-body reaction ($\rm H + H + H \rightarrow H_2 + H$). The
three-body reaction dominates at very high densities \citep[ $n >
  10^9$~cm$^{-3}$;][]{Galli1998}. For normal ISM conditions, H$_2$ is
predominantly formed on dust grains, and much progress has been made
over the past two decades, both experimentially as well as
theoretically, in understanding this (much faster) formation of H$_2$
on the surfaces of grains \citep[][and references
  therein]{pirronello1999,Katz1999,Cazaux2002,Zecho2002,Cazaux2004,Sha2005,Hornekaer2006,Morisset2004,Cuppen2010}. In
general, the understanding is that CO is formed in the gas-phase, but
the origin of gas-phase water is not so clear-cut. It is possible to
form water in the gas-phase through either a chain of neutral-neutral
reactions containing a number of temperature barriers, or through
ion-molecule reactions when the gas is moderately ionized ($x_e\sim
10^{-5}$). The formation of water on interstellar dust can also be an
efficient route in diffuse and dense clouds
\citep{Cuppen2007,Cuppen2010b}. Several routes to form water can be
considered, involving successive hydrogenation of oxygen either with
H$_2$ or H. Because species can stay on the dust, they can
repetitively collide with each other and react, even if there is an
important barrier for the reaction to occur. In this case, dust grains
provide a favorable place for improbable reactions (with high
barriers) to occur \citep{Cazaux2010}. Also, other routes to form
water, involving O$_2$ \citep{Miyauchi2008,Ioppolo2010} and O$_3$
\citep{Mokrane2009} have been investigated and can be important at
high dust temperatures ($T_{\rm dust}>20-30$~K). The molecules present
on the surface can be released in the gas phase through several
mechanisms: desorption upon formation if the reaction is exothermic;
photo-dissociation of surface species can lead to the desorption of
products in the gas \citep[i.e., photodissociation of H$_2$O leads to
  OH and H in the gas phase,][]{Andersson2006}; evaporation of the
surface when the dust temperature is high enough (this process depends
on the binding energy of the species). If many water molecules are
formed and stay on the dust, then ices can be made. The bonds that
hold the water molecules together are very strong (creating water
clusters).  As a result, when one or more monolayers of ice are formed
on the grain, shocks will be the only effective way to remove the ice
from grains \citep[e.g.,][]{Draine1995}. It is thus only possible to
have efficient formation of gas-phase water with grains as catalyst,
when the grains are (close to) bare.

Objects such as Mrk 231 have very violent environments in the central
regions around the super-massive black hole. This particular object
has a derived $8 - 1000~\mu$m infrared luminosity $L_{IR}=4.0\times
10^{12}$~L$_\odot$ \citep{Boksenberg1977}. A highly absorbed power-law
X-ray spectrum was observed by \citet{Braito2004}. It is in these
extreme environments, that we expect cloud-cloud collisions to cause C
and J shocks, and strip the grains of their ice layers. The internal
UV field created by X-rays from the accreting black hole does not
allow to reform the ice. The abundances of H$_2$O \citep[$x_{\rm
  H_2O}\sim 10^{-6}$,][]{Gonzalez2010} in the warm molecular gas of
Mrk 231 are high. Here we study the effect of OH and H$_2$O formation
on {\bf bare} dust grains in environments with strong X-ray radiation
fields, in order to investigate whether this process would be able to
significantly boost the total production rate of these species in
these environments. Our work differs from \citet{Hollenbach2009},
since we consider an additional process to release molecular species
into the gas phase, and also since our study only considers bare
interstellar dust grains. Even though this paper focuses on an
application to ULIRGs, it will have applications to other objects as
well, such as X-ray irradiated disks around Young Stellar Objects. The
manuscript is articulated in the following way: We first derive
analytical expressions for the formation rates of OH and H$_2$O on
dust grains. Then we compare these rates to the neutral-neutral
gas-phase formation processes. Finally, the effects of the grains
surface chemistry on the XDR thermo-chemical structure are shown.
Their implications to explain recent observations with the {\it
  Herschel} Space Observatory are also discussed.

\section{H$_2$O and OH formation rates}

As mentioned earlier, gas-phase OH and H$_2$O can occur through
neutral-neutral reactions, or when the medium is moderately ionized,
$x_e \gtrsim 10^{-5}$, through ion-molecule reactions. The
neutral-neutral reactions OH + H$_2$ $\rightarrow$ H$_2$O + H and O +
H$_2$ $\rightarrow$ OH + H are slow at low temperatures, because they
have activation barriers \citep{Wagner1987}, and only contribute at
temperatures $T > 250-300$~K). The other route is H$^+$ + O
$\rightarrow$ O$^+$ + H, O$^+$ + H$_2$ $\rightarrow$ OH$^+$ + H,
OH$^+$ + H$_2$ $\rightarrow$ H$_2$O$^+$ + H, H$_2$O$^+$ + H$_2$
$\rightarrow$ H$_3$O$^+$ + H, followed by recombination H$_3$O$^+$ +
e$^-$ $\rightarrow$ H$_2$O + H.

In our current X-ray chemistry code, only the grain surface reaction
is taken into account for the formation of H$_2$. In order to study
the effect of OH and H$_2$O molecule formation on grain surfaces,
similar reactions for these species will be added to the chemical
network 

\begin{eqnarray}
\frac{dn({\rm OH})}{dt} &=& n({\rm O}) v_{\rm O} n_{\rm dust} \sigma
\epsilon_{\rm OH}, \\
\frac{dn({\rm H_2O})}{dt} &=& n({\rm O}) v_{\rm O} n_{\rm dust} \sigma
\epsilon_{\rm H_2O},
\end{eqnarray}

\noindent
where $\epsilon_{\rm OH}$ and $\epsilon_{\rm H_2O}$ are efficiencies
for the formation of OH and H$_2$O, respectively, and $v_{\rm O}$ the
velocity of oxygen atoms. The expression for the efficiencies will be
determined in the next section, where we calculate what fraction of an
accreted oxygen atom is converted in gas phase OH or H$_2$O. Two
different grain size distributions are considered, an MRN
\citep{Mathis1977} distribution and a \citet{Weingartner2001}
distribution, with $n_{\rm dust}\sigma / n({\rm H}) =
10^{-21}$~cm$^{-2}$ and $2.8\cdot 10^{-21}$~cm$^{-2}$,
respectively. The MRN distribution considers a minimum dust grain size
of 50\AA, while Weingartner and Draine (W\&D) consider much smaller
grains, which are actually PAHs, and can be as small as 5\AA. The W\&D
grain size distribution, however, seems more appropriate to
environments such as the central regions of Mrk231, where dust grains
can break in smaller particles due to, e.g., shocks. A larger cross
section yield more enrichment of the medium with water and OH. The
efficiencies are derived in the next section.

\section{Grain surface chemistry}

In this study, we consider carbon grains (graphite and
carbonaceous). In order to describe the population of a certain
species on a grain, we need to consider additional processes apart
from the reactions that normally occur in the gas phase, such as
neutral-neutral reactions. These are: accretion to, evaporation from,
tunneling and thermal hopping of atoms and molecules on the surface of
dust grains.

Species arriving from the gas phase arrive at a random time and
location on the dust surface. The arrival time depends on the
accretion rate, which can be written as

\begin{equation}
R_{acc} = n_X v_X \sigma S_{X}\label{accretion_rate}
\end{equation}

\noindent
where $n_X$ and $v_X$ are the densities and velocities, respectively,
of the species $X$, and $\sigma$ is the cross-section of the dust
particle. $S_{X}$ is the sticking coefficient of the species with the
dust, and we consider $S_{X}=1$. The species $X$ can go back to the
gas phase through evaporation. This rate can be written as

\begin{equation}
R_{evap} = \nu_i \exp(-E_X / kT)\label{evaporation_rate}
\end{equation}

\begin{table}
\caption{Binding energies (in K) for carbon grains}   
\label{binding_energy}    
\centering         
\begin{tabular}{l l l}    
\hline\hline                
\vspace{-1.0mm}    \\
Physisorption
\vspace{1mm}    \\
\hline 
\vspace{-1mm} \\    
H      &  550           & \citet{Bergeron2008} \\
O      &  1390          & \citet{Bergeron2008} \\
H$_2$  &  600           & \citet{Akai2003} \\
OH     &  1360          & \citet{Cuppen2007}\\
O$_2$  &  1440          & \citet{Cuppen2007} \\
H$_2$O &  2000          & \citet{Cuppen2007} \\
O$_3$      & 2240       & \citet{Lee2009} \\
HO$_2$     & 2160       & \citet{Cuppen2007}\\
H$_2$O$_2$ & 2240       & \citet{Cuppen2007}\\
\vspace{-1mm}    \\
\hline 
\vspace{-1mm} \\     
Chemisorption:
\vspace{1mm}    \\
\hline 
\vspace{-1mm} \\   
H       &  8500         &  \citet{Jelea2004} \\
\vspace{-1mm}    \\
\hline                           
\end{tabular}
\end{table}

\noindent
with $\nu_i$ is the oscillator factor of atom in site $i$ (which is of
order 10$^{12}$ s$^{-1}$ for physisorbed atoms), and $E_X$ the binding
energy of species $X$ (see Table \ref{binding_energy} for the
  adopted values). The binding energy $E_X$ can be either weak (for
atoms and molecules) or strong (only for atoms). In this study we
consider binding energies similar to those in
\citet{Cazaux2010}. There are two types of interactions between the
atoms and the surface: a weak one, called physisorption (van der Waals
interaction), and a strong one, called chemisorption (covalent
bound). Atoms from the gas phase can easily access the physisorbed
sites and become physisorbed atoms. These weakly bound atoms can
travel on the surface at very low dust temperatures, meeting each
other to form molecules. Moving of species on dust grains occcurs
through tunneling or thermal hopping \citep{Cazaux2004}. The
  mobility of a species $X$ moving from physisorbed sites to
  physisorbed sites can be written as:

\begin{equation}
\alpha_{pp}(X)=\nu P_{pp}(X),\label{hopping_rate}
\end{equation}

\noindent
where $P_{pp}(X)$ is the probability for the atom to move from a
physisorbed to another physisorbed site. In this study, we consider a
range of surface temperatures $T > 15$~K. At these temperatures,
thermal hopping dominates and the probability for a species $X$ to go
from a physisorbed site to another physisorbed site can be written as:

\begin{equation}
P_{pp}(X) = \exp(-E_{pp}(X) / kT),
\end{equation}

\noindent
where $E_{pp}(X)$ is the barrier between to physisorbed sites, that we
assume to be 2/3 of the binding energy $E_X$ (see Table
\ref{binding_energy}). Hydrogen atoms that are physisorbed can also
enter chemisorbed sites through tunneling effects or thermal
hopping. Because of the high barrier against chemisorption
\citet{Sha2002}, H atoms usually tunnel through the barrier to
populate chemisorbed site. The mobility for the H atoms can be written
as:

\begin{equation}
 \alpha_{pc}=\nu P_{pc},
\end{equation}

\noindent
where $P_{pc}({\rm H})$ is the probability for a physisorbed H atom to
enter a chemisorbed site \citep{Cazaux2004}. Once two species meet in
the same site, they can form a new species if the activation barrier
for the formation can be overcome. According to the exothermicity of
the reaction, the newly formed species is {\it directly released into
  the gas phase}. This probability, called $f_{\rm{des}}$ in this
work, has been discussed in \citet{Cazaux2010}, and is the most
important route to release products to the gas-phase. Species present
on the surface can also receive a UV photon and become
photo-dissociated. In X-ray exposed environments, an internal UV field
is created by excitation of Lyman-Werner bands, and about 30 percent
of the locally absorbed X-ray is converted into UV
\citep{Maloney1996}. The photo-dissociation of H$_2$O on the dust can
lead to the release in the gas phase of the two products of the
dissociation, OH and H. This mechanism has a very small chance to
occur \citep[2 percent,][]{Andersson2008}. Although this process is
much faster than thermal desorption (the grains are too cold to make
thermal desorption efficient), it is not the main contributor in
delivering species to the gas-phase. On grains, where ice layers are
present, the desorption upon formation is less prominent, because the
species are more strongly bound to the surface. In those environments,
photo-desorption is the dominant route to release species into the gas
phase. More details about the adopted reaction rates can be found in
\citet{Cazaux2010}. As we are interested in the oxygen chemistry, we
will consider H, H$_2$, O, OH, O$_2$, H$_2$O, O$_3$, HO$_2$, and
H$_2$O$_2$ as grain surface species in the chemical network.

Two different numerical methods are performed to follow the formation
of species on dust grain surfaces: (1) a Monte Carlo and (2) a rate
equations method. MC simulations are performed at several fixed dust
temperatures. These calculations are very time consuming, and are used
in order to check for which conditions the chemistry occuring on the
dust enter the stochastic regime. The rate equations method is used to
simulate the chemistry occuring onto dust for a wide range of dust and
gas temperatures, gas phase chemical composition, radiation
fields. This method is orders of magnitude faster than MC simulations.

\subsection{Monte Carlo method}

\begin{figure*}
  \centering
  \includegraphics[width=9cm,clip]{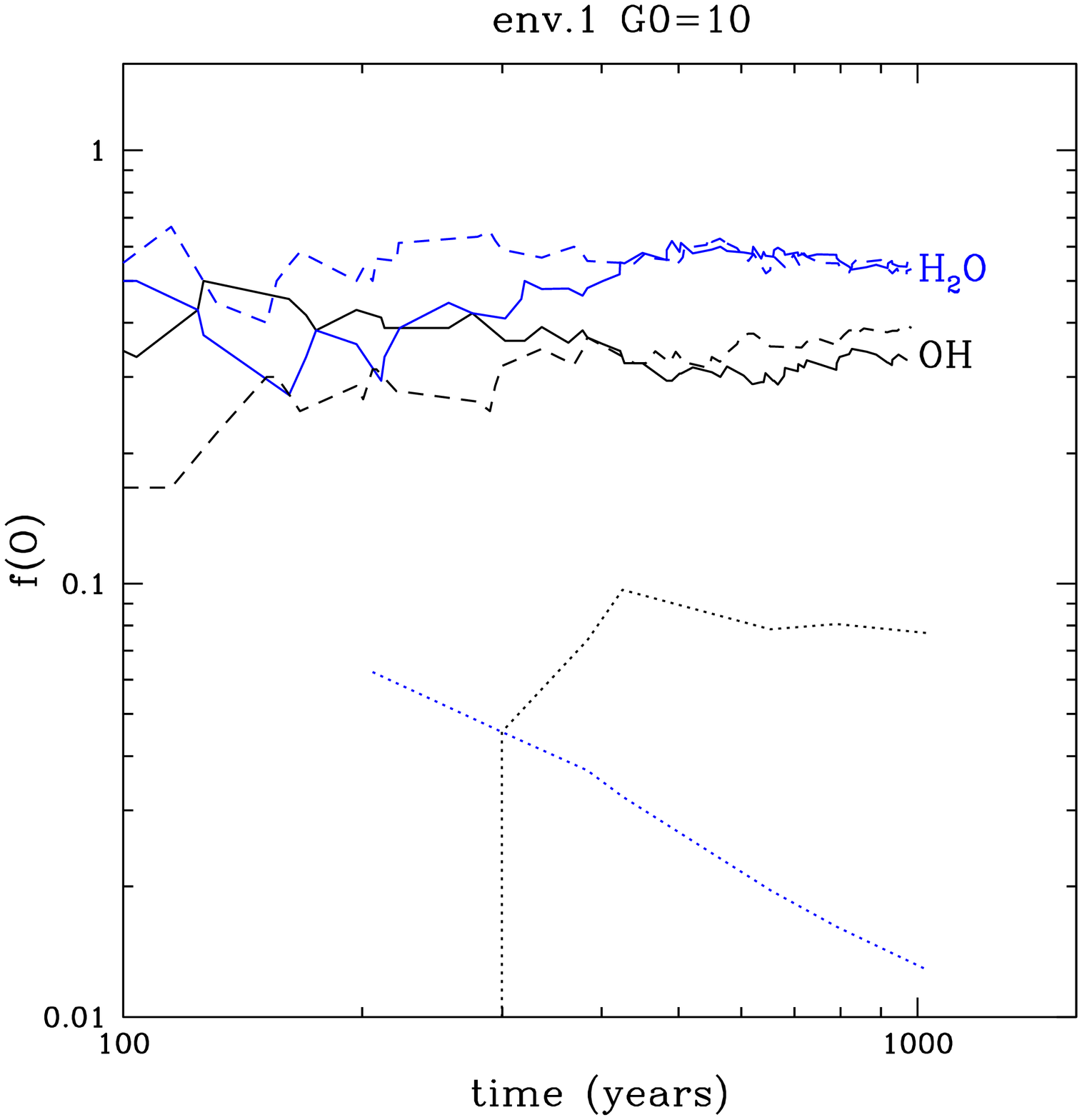}
  \includegraphics[width=9cm,clip]{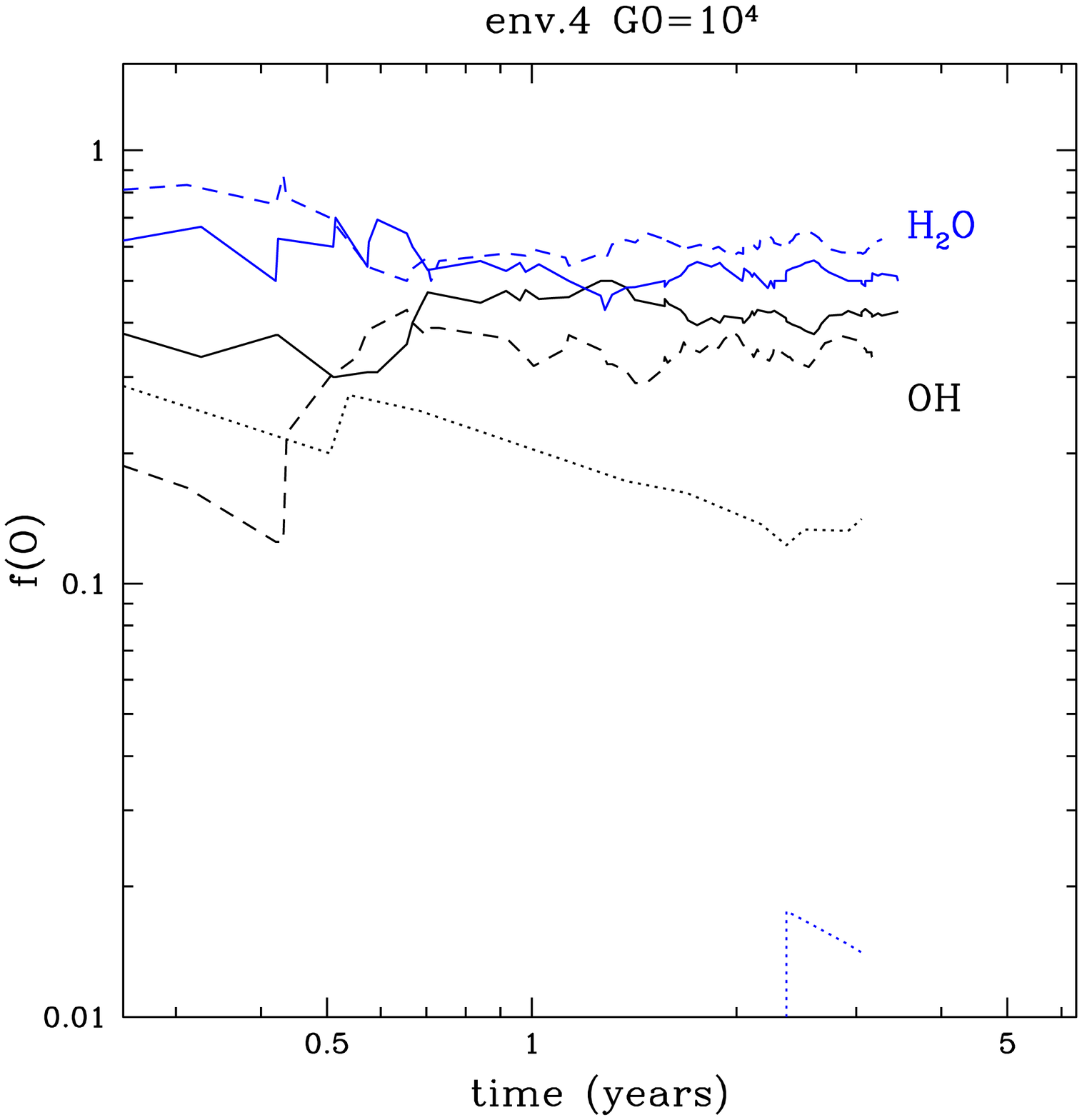}
  \includegraphics[width=9cm,clip]{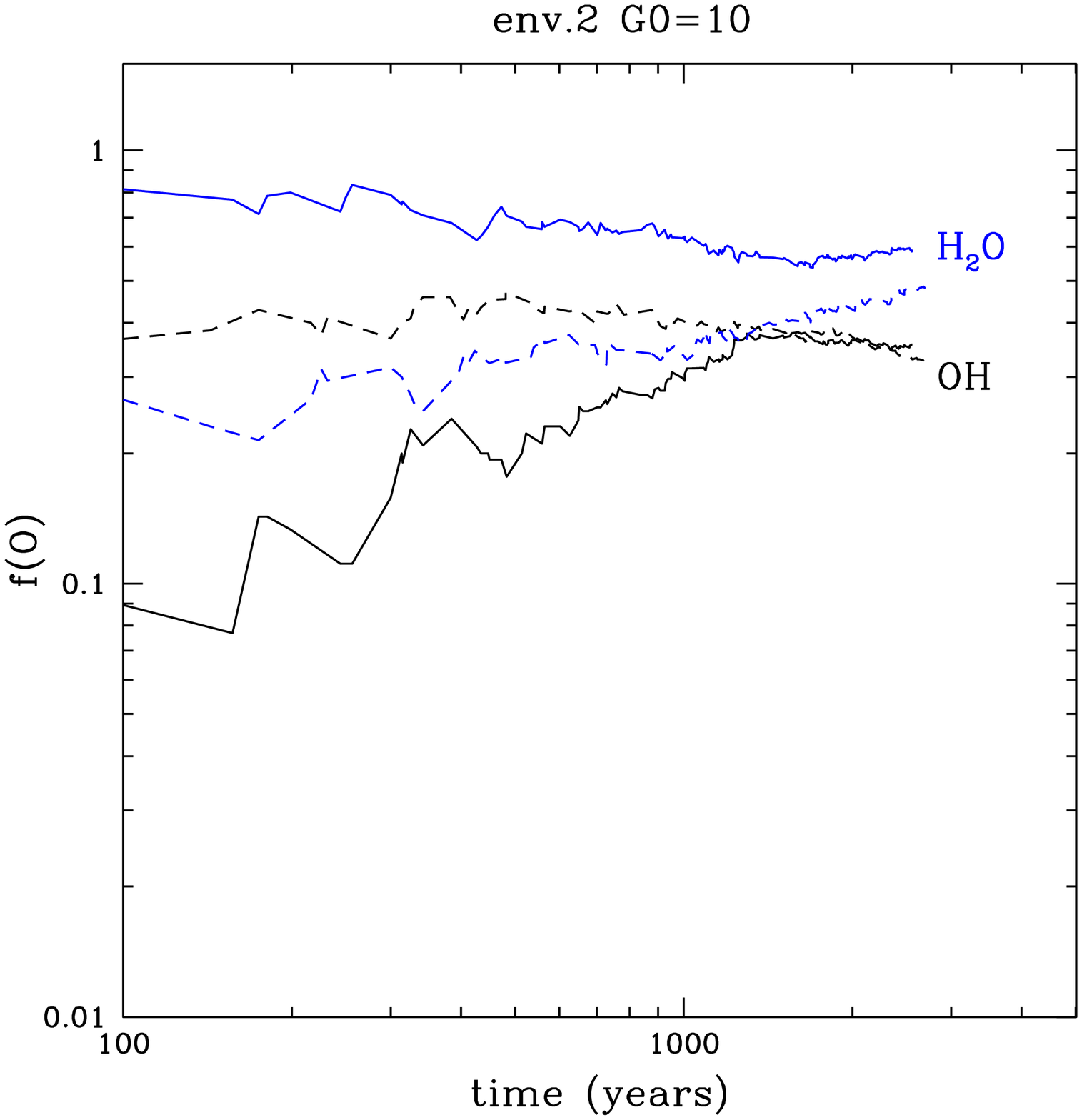}
  \includegraphics[width=9cm,clip]{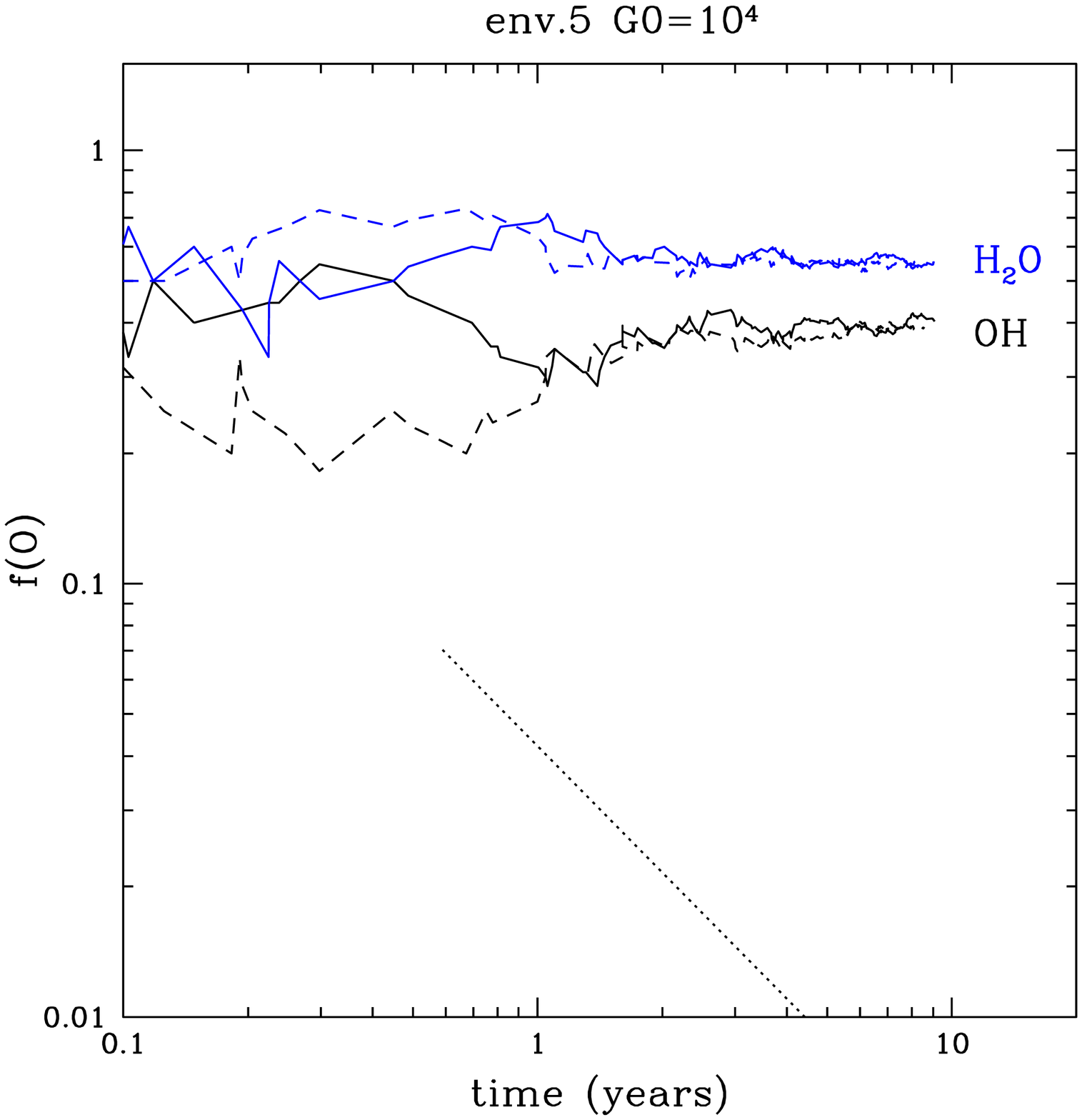}
  \caption{Results of our Monte Carlo simulations on a 10$\times$10 sites grain (30~\AA) for environments 1,2, 4 and 5. The solid lines represent our results for $T_{\rm dust}=20$~K, dashed lines for $T_{\rm{dust}}=30$~K and dotted lines $T_{\rm{dust}}= 35$~K (left panel) 40~K (right panel). }
  \label{MC}
\end{figure*}

We use step-by-step Monte Carlo simulations to follow the chemistry
occurring on dust grains. The dust grains are divided into square
lattices of 10$\times$10 adsorption sites. Each site on the grid
corresponds to a physisorbed site and a chemisorbed site, so that the
grain can be seen as two superimposed grids. Species originating from
the gas phase arrive at a random time and location on the dust
surface. This arrival time depends on the rate at which gas species
collide with the grain (Eq. \ref{accretion_rate}). When a species
arrives at a point of the grid, it can become physisorbed, or, if its
chemisorption states exists and its energy is high enough to cross the
barrier against chemisorption, it becomes chemisorbed. In our model,
however, because of the high barrier to access chemisorbed sites, H
atoms mostly arrive from the gas phase in physisorbed sites. The
species present on the surface may return to the gas phase if they
evaporate (Eq. \ref{evaporation_rate}). The species that arrive at a
location on the surface can move randomly across the surface by means
of tunneling effects and thermal hopping
(Eq. \ref{hopping_rate}). Species present on the surface can also
receive a UV photon and become dissociated at the rates shown in table
\ref{photorates}. Once a chemical species present on the dust moves,
or adsorbs a photon, or meets another species, the next event that
occurs to this species is determined as well as its time of
occurrence. Therefore, the events that concern every species on the
dust are ordered by time of occurrence, and for each event that
occurs, a next event for the concerned species is determined. In
figure ~\ref{MC}, we follow the fraction of oxygen that is released in
the gas under the form of OH and H$_2$O for environments 1, 2, 4 and 5
(see Table \ref{params_env} for adopted parameters). The parameters of
these environments represent typical ambient low and high densities
and radiation fields in the ISM of active galaxy centers, and are
similar to those in \citet{Meijerink2005}. The solid and dashed lines
show our results for $T_{\rm{dust}}$ = 20 and 30 K respectively.  For
each environments, 50-60 $\%$ (30-40$\%$) of the oxygen arriving on
the dust goes back in the gas phase under the form of H$_2$O (OH). At
higher dust temperatures ($\sim 35-40$~K), on the other hand, species
coming on the dust evaporate very fast. Therefore, the efficiencies
decrease dramatically since the grain surfaces are only sporadically
covered by species.

\subsection{Rate equations method}

The rate equations method can be applied when there is a minimum
coverage of 1 atom on the dust. This condition is satisfied if the
grains are not too warm, that species can encounter each other before
evaporating. By comparing our rate equations results with Monte Carlo
simulations, we will define a range of dust temperatures for which
rate equations are valid.

The rate equation method couples the different equations for the
different species on the dust. The processes described in the previous
subsection are dependent on the dust and gas temperature, UV field,
and exo-thermicity of the reaction occurring on the dust. The equation
for species $X$ can thus be written as follows:

\begin{eqnarray}
\frac{dn(X)}{dt} &=&n_{\rm gas}(X) v_X N_s-n(X) n(Y) \alpha_{pp}(X)+  \nonumber \\
                 & & (1-f_{des}) n(Y) n(Z) \alpha_{pp}(Y)  -R_{evap_X} n(X) \nonumber \\
                 & &-R_{phot_X} n(X)+R_{phot_Y} n(Y)
\end{eqnarray}

\noindent where $\alpha_{pp}(X)$ and $\alpha_{pp}(Y)$ are the
mobilities of the species $X$ and $Y$ (see Eq.
\ref{hopping_rate}). The different terms of this equation present (1)
accretion of species $X$ from the gas phase, (2) the formation of the
new species that involve species $X$, (3) creation of the species $X$
by the encountering of species $Y$ and $Z$ that stay on the surface,
(4) evaporation of species $X$, (5) photo-dissociation of $X$ on the
dust, and (6) the creation of $X$ by photo-dissociation of species
$Y$.

\begin{table}
\caption{Models parameters}   
\label{params_env}    
\centering         
\begin{tabular}{l c c c c c}    
\hline\hline                
\vspace{-1mm}    \\
Env.          & $n_{\rm H}$ & $x({\rm HI})$ &  $x({\rm H_2})$ & $x({\rm OI})$ & $G_0$ \\
\vspace{-1mm}    \\
\hline 
\vspace{-1mm} \\       
1 & $10^3$    & $1$     & $10^{-3}$     & $3.4\times 10^{-4}$ & $10^0$,$10^1$,$10^2$,$10^3$ \\
2 &   ,,      & $10^{-1}$ & 0.45         & ,,           & ,,\\
3 &   ,,      & $10^{-2}$ & 0.5         & ,,           & ,,\\
4 & $10^{5.5}$ & $1$ & $10^{-3}$  & ,,            &  $10^3$,$10^4$,$10^5$,$10^6$ \\
5 &   ,,        & $10^{-1}$ & 0.45        & ,,            & ,,\\
6 &   ,,        & $10^{-2}$ & 0.5        & ,,            & ,,\\
\vspace{1mm}    \\
\hline                           
\end{tabular}
\end{table}

\subsection{Efficiencies derived from rate equation method}

\begin{figure*}
  \centering
  \includegraphics[width=16cm,clip]{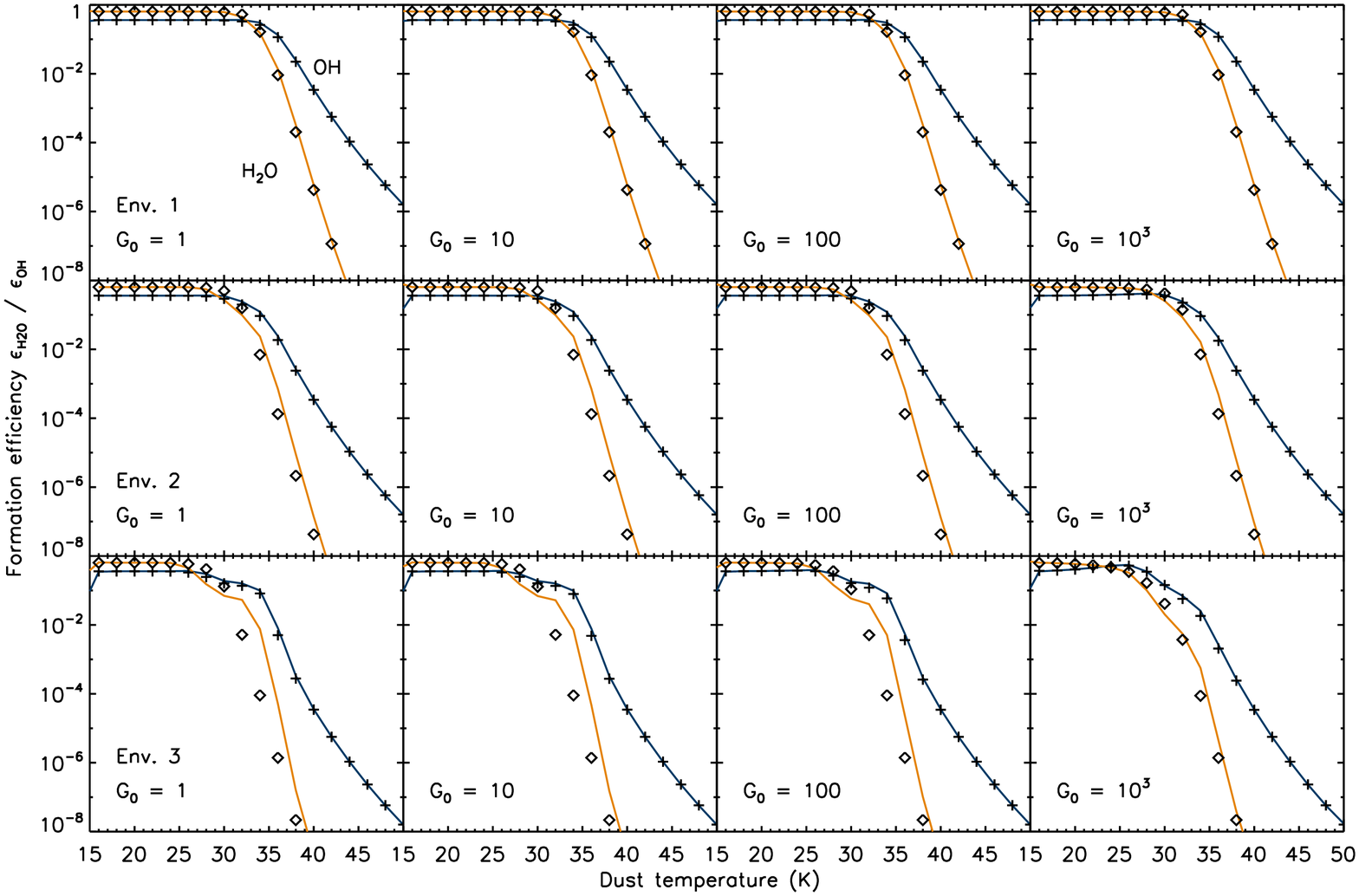}
  \caption{Efficiencies for the formation of OH and H$_2$O in low
    density gas ($n=10^3$~cm$^{-3}$), in atomic and molecular hydrogen
    dominated gas. UV fields ranging from $G_0$=1 to 1000 are
    considered. Results from rate equation are shown by a solid and a
    dotted line for OH and H$_2$O, respectively. The analytical fit is
    shown by the crosses (OH) and diamonds (H$_2$O).}
  \label{low_dens_eff}
\end{figure*}

\begin{figure*}
  \centering
  \includegraphics[width=16cm,clip]{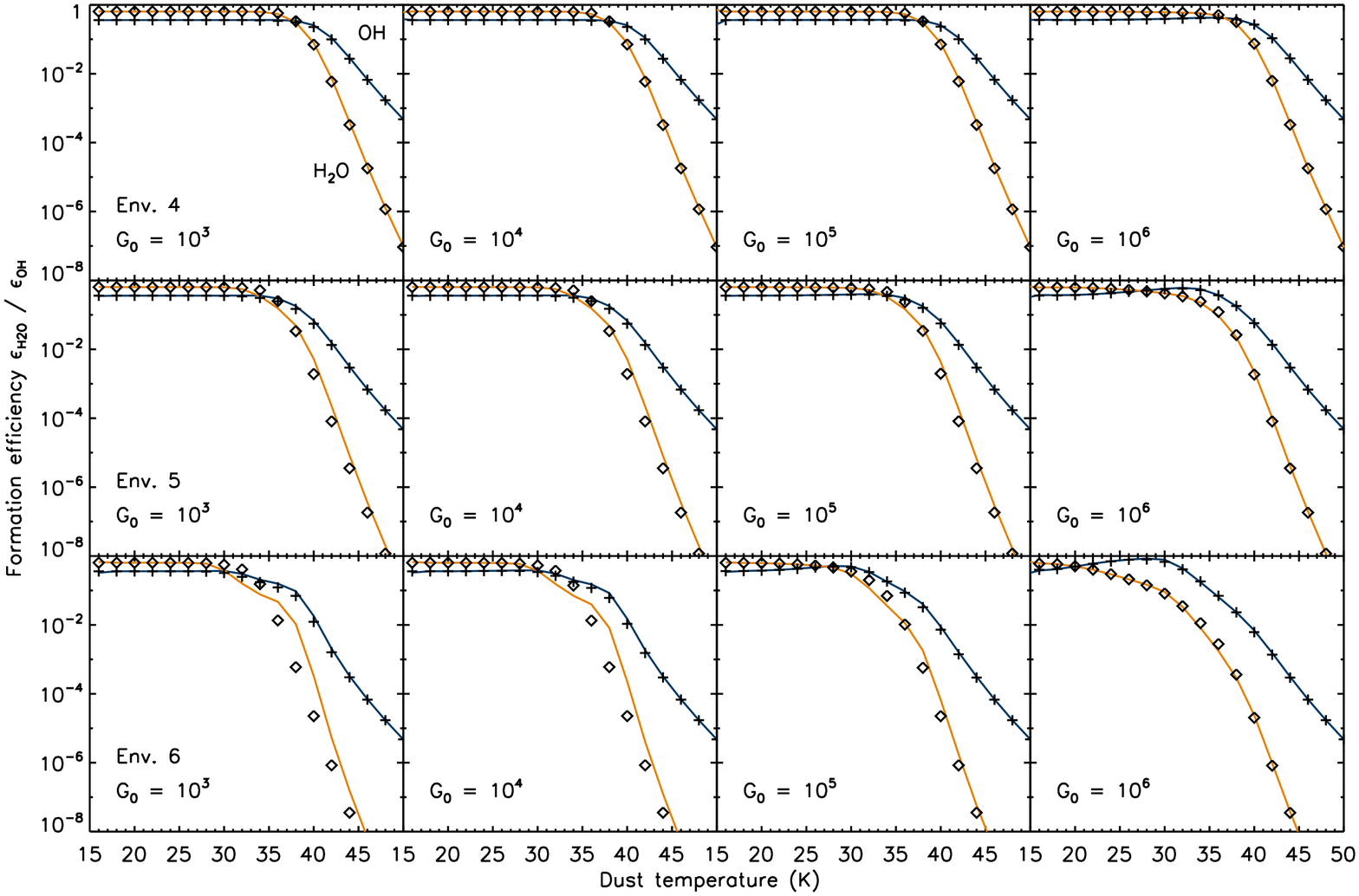}
  \caption{Efficiencies for the formation of OH and H$_2$O in high
    density gas ($n=10^{5.5}$~cm$^{-3}$), in atomic and molecular
    dominated gas. UV fields ranging from $G_0=10^3$ to $10^6$ are
    considered. Results from rate equation are shown by a solid and a
    dotted line for OH and H$_2$O, respectively. The analytical fit is
    shown by the crosses (OH) and diamonds (H$_2$O).}
  \label{high_dens_eff}
\end{figure*}

\begin{figure*}
  \centering
  \includegraphics[width=16cm,clip]{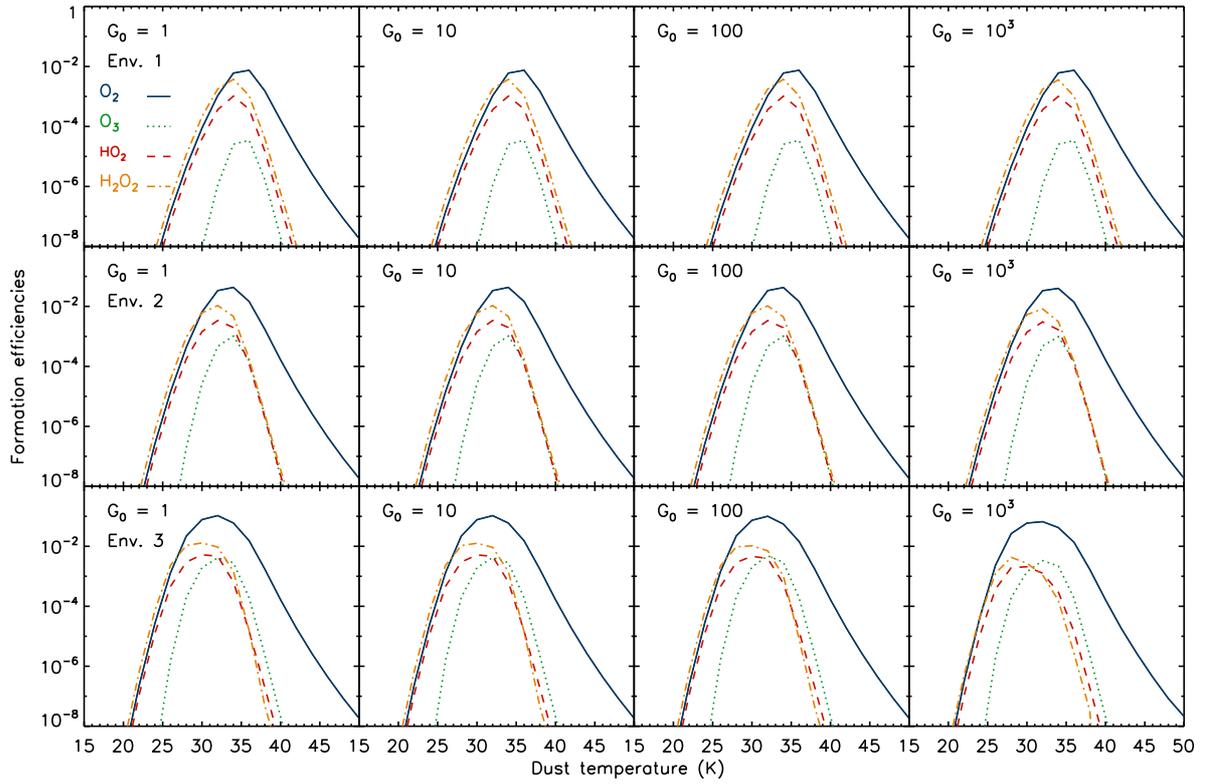}
  \caption{Efficiencies for the formation of oxygen bearing species
    O$_2$, O$_3$, OH$_2$, and H$_2$O$_2$ in low density gas
    ($n=10^3$~cm$^{-3}$), in atomic and molecular hydrogen dominated
    gas. UV fields ranging from $G_0$=1 to 1000 are
    considered.}
  \label{low_dens_eff_others}
\end{figure*}

\begin{figure*}
  \centering
  \includegraphics[width=16cm,clip]{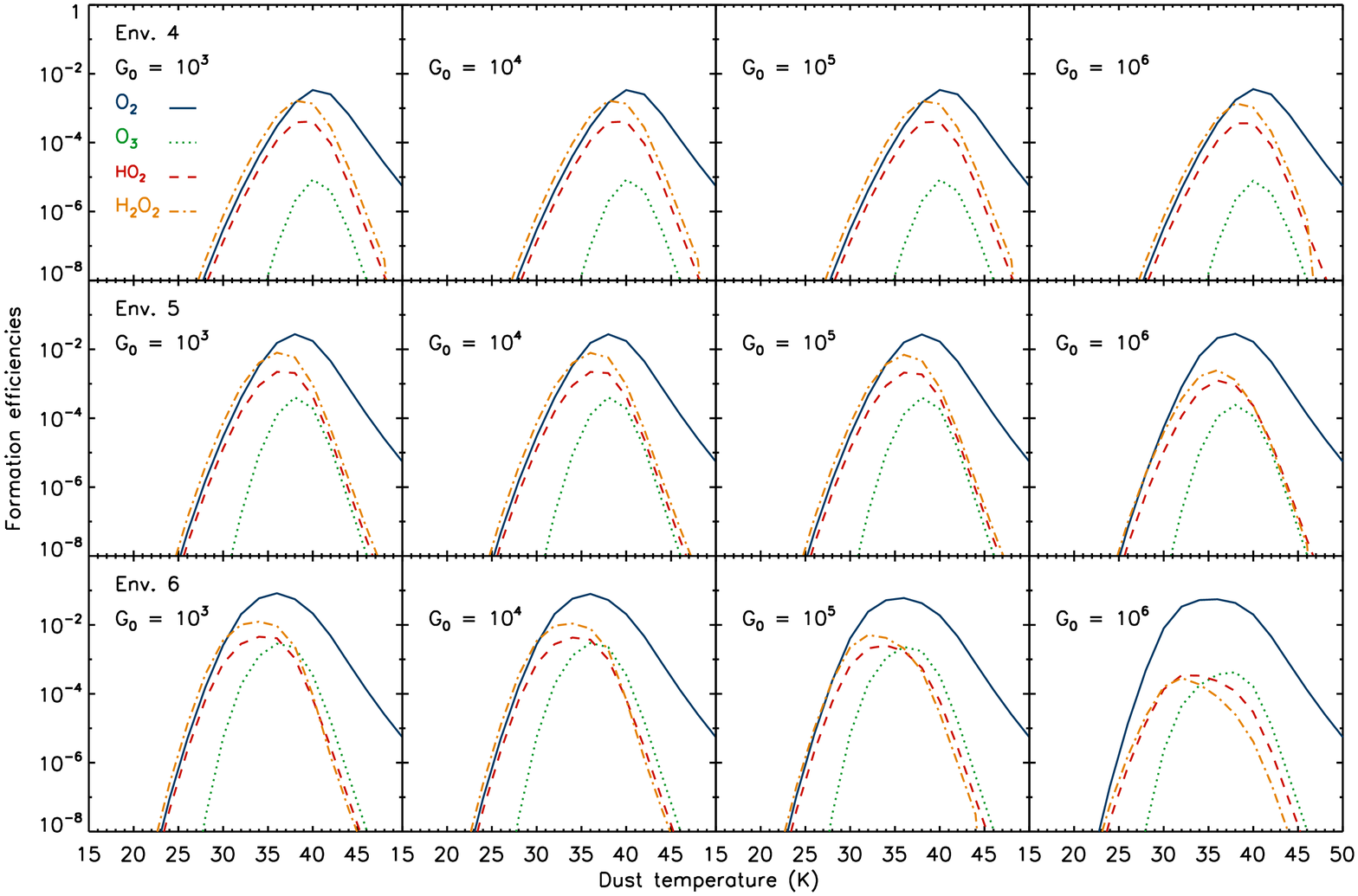}
  \caption{Efficiencies for the formation of oxygen bearing species
    O$_2$, O$_3$, OH$_2$, and H$_2$O$_2$ in high density gas
    ($n=10^{5.5}$~cm$^{-3}$), in atomic and molecular dominated
    gas. UV fields ranging from $G_0=10^3$ to $10^6$ are considered.}
  \label{high_dens_eff_others}
\end{figure*}

The formation efficiency is defined as the fraction of oxygen $f({\rm
  O})$ that is released in the gas phase under the form of another
species. We studied this in a few different environments, chosen such
that they resemble the conditions as found in the XDR models of
Meijerink \& Spaans (2005). The gas temperature is fixed at a constant
temperature $T=100$~K, and dust temperatures ranging from
$T_{\rm{dust}}=10-60$~K. Four different radiation fields were
considered at two fixed densities, $n=10^3$ and $10^{5.5}$~cm$^{-3}$,
in both the atomic and molecular regime. The parameters used are
summarized in Table \ref{params_env}.

In Figs. \ref{low_dens_eff} and \ref{high_dens_eff}, we show the
obtained efficiencies for OH and H$_2$O. The figures show that between
dust temperatures $T_{\rm dust}\sim 15-35$~K, the efficiencies for OH
and H$_2$O are roughly constant. In this regime most of the oxygen
goes to H$_2$O. For dust temperatures $T_{\rm{dust}} \gtrsim 35-45$~K,
the total efficiency for forming molecules is rapidly decreasing, and
most of the atomic oxygen will not react to another species, before it
evaporates from the grain. Therefore, the dust surface chemistry can
only be effective, when the dust temperatures are not too high. Recall
that the way to form OH and H$_2$O is through $\rm O + H \rightarrow
OH$ and $\rm OH + H \rightarrow H_2O$. For slightly higher dust
temperatures, the chance of an OH molecule to evaporate from the grain
becomes higher than that for meeting another H atom. Therefore at
temperatures $T_{\rm{dust}} \gtrsim 35 - 40$~K, most of the oxygen
atoms are incorporated into OH, and then $\epsilon_{\rm OH}$ is
highest.

Increasing the radiation field, has a similar effect as increasing the
dust temperature, i.e., it results in a lower $\epsilon_{\rm H_2O}/
\epsilon_{\rm OH}$ ratio, which is caused by photo-dissociation
  of H$_2$O on the surface. The strongest effect is seen in the
environments 3 and 6, where the dominant hydrogen fraction is
contained in H$_2$.  In this environment, the hydrogen accretion rates
are lowest, and thus also have the lowest atomic surface density on
grains. When most of the gas is in atomic form, the hydrogen accretion
is such that at high dust temperatures most of the oxygen is converted
into OH. However, when the gas become more and more molecular, the
atomic hydrogen accretion rate onto the dust decreases, and a larger
fraction of atomic oxygen ends up in O$_2$. In environment 3, this is
even the most likely product at temperature $T_{\rm{dust}} > 35$~K.

In Figs. \ref{low_dens_eff_others} and \ref{high_dens_eff_others}, the
efficiencies for the formation of O$_2$, O$_3$, OH$_2$, and H$_2$O$_2$
are shown. The efficiencies for these species show a peak at dust
temperatures $T_{\rm dust}=35-40$~K. The maximum efficiency for
producing O$_2$ ranges from $\sim 10^{-2}$ in the atomic gas to $\sim
10^{-1}$ in molecular gas, and generally the efficiencies for other
species are smaller. At temperatures $T > 40$~K, O$_2$ efficiencies
dominate over those for OH, but this is the regime where efficiencies
for the formation of molecular species become very small.

\subsection{Validity of rate equation method}

In order to estimate the uncertainties in the MC calculation, and to
check for consistancy between the MC and rate equation method, we did
two additional MC simulations at dust temperatures of 20, 30, and 35
(low density) and 40 (high density). After a time $t$ of $\sim 1000 -
2000$~years (low density) and $\sim 5 - 10$~years (high density), the
efficiencies are converged and oscillate around a stable value.  At
$T_{\rm dust} = 20$ and 30~K, the H$_2$O efficiencies vary between 0.5
to 0.6, and those for OH between 0.3 to 0.4, and thus very similar to
the rate equation method. At the higher dust temperatures (35 and
40~K), the values obtained in the MC method are lower. The rate
equations method is overestimating the efficiencies, and we are
entering the stochastic regime.

We only concentrate on dust temperature $T_{\rm dust} > 10 - 15$~K. At
very low dust temperatures, OH and H2O are still able to form at a
smaller efficiency, but the chemistry is different then for a number
of reasons: (1) ices can form on the surface (2) the chemistry is
different, because the binding energies are different, and (3) the
surface of the grain will be saturated with H$_2$.

\section{Analytical expressions for OH and H$_2$O formation efficiencies}

The rate equations method gives the populations of the different
species on the dust surface, which allows us to derive simplified
analytic expression for formation efficiencies of OH and H$_2$O, that
is easy to use in numerical codes. This means that the smallest set of
reactions needed to reproduce the derived efficiencies from the rate
equation calculation is determined. Although the analytic expression
of the efficiencies is derived in Appendix \ref{deriv_eff}, the
dependencies are given here:

\begin{eqnarray}
\epsilon_{\rm OH} &=& \frac{f_{des_{\rm OH}} n({\rm O}) n(\rm H) \alpha_{\rm H}}{R_{acc}({\rm O})} +
\frac{f_{des_{\rm OH'}} n({\rm O_3}) n({\rm H}) \alpha_{\rm H}}{R_{acc}({\rm O})}\label{OHeff_final} \\
\epsilon_{\rm H_2O} &=& \frac{f_{des_{\rm H_2O}} n({\rm OH}) n(\rm H) \alpha_{\rm H}}{R_{acc}({\rm O})}.  \label{H2Oeff_final}
\end{eqnarray}

The fractions $f_{des_{\rm OH}}$, $f_{des_{\rm OH'}}$, and
$f_{des_{\rm H_2O}}$ are 0.36, 0.25, and 0.15, respectively. These are
the fractions of the molecules that are released in the gas phase upon
formation due to the exothermicity of the reaction. $R_{acc}(\rm O)$
is the accretion rate of oxygen, while $n(\rm O)$, $n(\rm O_3)$, and
$n(\rm OH)$ are the dust surface densities of O, O$_3$, and OH,
respectively, in monolayers, e.g., 1 monolayer = 100\% coverage. These
surface densities are listed in Appendix \ref{deriv_eff}. $\alpha_{\rm
  H}$ is the mobility of hydrogen atoms on the grain surface. The
efficiency for the formation of water does {\it not} contain a term
with accretion of OH from the gas phase, since the accretion of O
followed by OH formation dominates over OH accretion.

\begin{figure}
  \centering
  \includegraphics[width=8cm,clip]{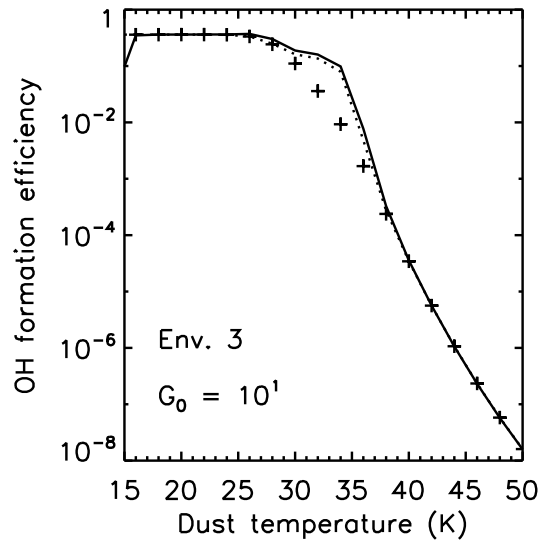}
  \caption{Efficiencies for environment 3, largely molecular, low
    density $n=10^3$~cm$^{-3}$ gas, $G_0=10^4$: Result from rate
    equations simulation (solid line), with (dotted line) and without
    (crosses) inclusion of the reaction $\rm O_3 + H \rightarrow OH +
    O_2$ in the analytical expression.}
  \label{O3_importance}
\end{figure}

A very important outcome of the simulation is that the formation of OH
on grains is dominated by $\rm O + H \rightarrow OH$ up to
temperatures $T \sim 30 - 40$~K depending on the environment, and by
$\rm O_3 + H \rightarrow OH + O_2$ at higher temperatures. This
implies that apart from expressions for $n(\rm O)$ and $n(\rm OH)$,
also those for $n(\rm O_2)$ and $n(\rm O_3)$ are needed. For
completeness, we also give the one for $n(\rm H_2O)$. The importance
of the $\rm O_3 + H$ reactions is shown in
Fig. \ref{O3_importance}. It is obvious from the figure that the OH
formation efficiency would be underestimated without the consideration
of the reaction with $\rm O_3$. These results are not inconsistent
with those obtained by \citet{Ioppolo2008}, where the formation of
H$_2$O is dominated by the reaction sequence: $\rm H + O_2 \rightarrow
HO_2$, $\rm H + HO_2 \rightarrow H_2O_2$ followed by $\rm H + H_2O_2
\rightarrow H_2O + OH$. We do consider the same reaction rate paths in
our work, but in this study, molecules are formed on {\it bare
  grains}, while \citet{Ioppolo2008} consider {\it icy grains
  mantles}. In our simulation a very significant part of the molecules
are desorbed upon formation, while the cycle of photodissociation of
H$_2$O and reformation on the grains significantly enhances the amount
of water released in the gas phase. {\it It does not require the
  complex route through $\rm H_2O_2$}. There is also another way to
form H$_2$O that involves O$_3$. The reaction $\rm O + O_2 \rightarrow
O_3$, that releases only a very small part into the gas phase upon
formation (only a few percent). So although the efficiency for the
formation of $\rm O_3$ is low, there is a significant amount of $\rm
O_3$ on the surface to react to OH. The work by \citet{Hollenbach2009}
is also very different from ours. They also treat a surface chemistry
network, with H$_2$, OH, O$_2$, and some other different species, such
as CH, CH$_2$, and CH$_3$. They do not include processes with O$_3$
and H$_2$O$_2$, intermediate species in the processes leading to water
formation. However, the biggest difference between our models are the
{\it dominant} processeses to deliver species back into the gas-phase
which are thermal, photo, and cosmic-ray desorption in the
\citet{Hollenbach2009} paper, as they are considering ices in that
work, whereas this is desorption upon formation in our work. They
consider a PDR environment, and therefore they allow for the formation
of strongly bound ice-layers. We assume that this does not occur
because of the turbulent motions stripping the grains and the strong
photodissociating radiation fields that are internally created by
X-rays.

The approximate analytical fits to the rate equation results are very
good for environments 1, 2, 4 and 5. However, environment 3 and 6,
which are highly molecular, start to show deviations for dust
temperatures $T_{\rm dust} > 30$~K for the H$_2$O. This is due to
additional reactions on the grain surface, that we did include in the
rate equation method, but not in the approximate fit.

The fits should not be applied at very low dust temperatures ($T_{\rm
  dust}) < 10$~K, because the analytical fits do not match the MC and
rate equation results.

\begin{figure*}
  \centering
  \includegraphics[width=8cm,clip]{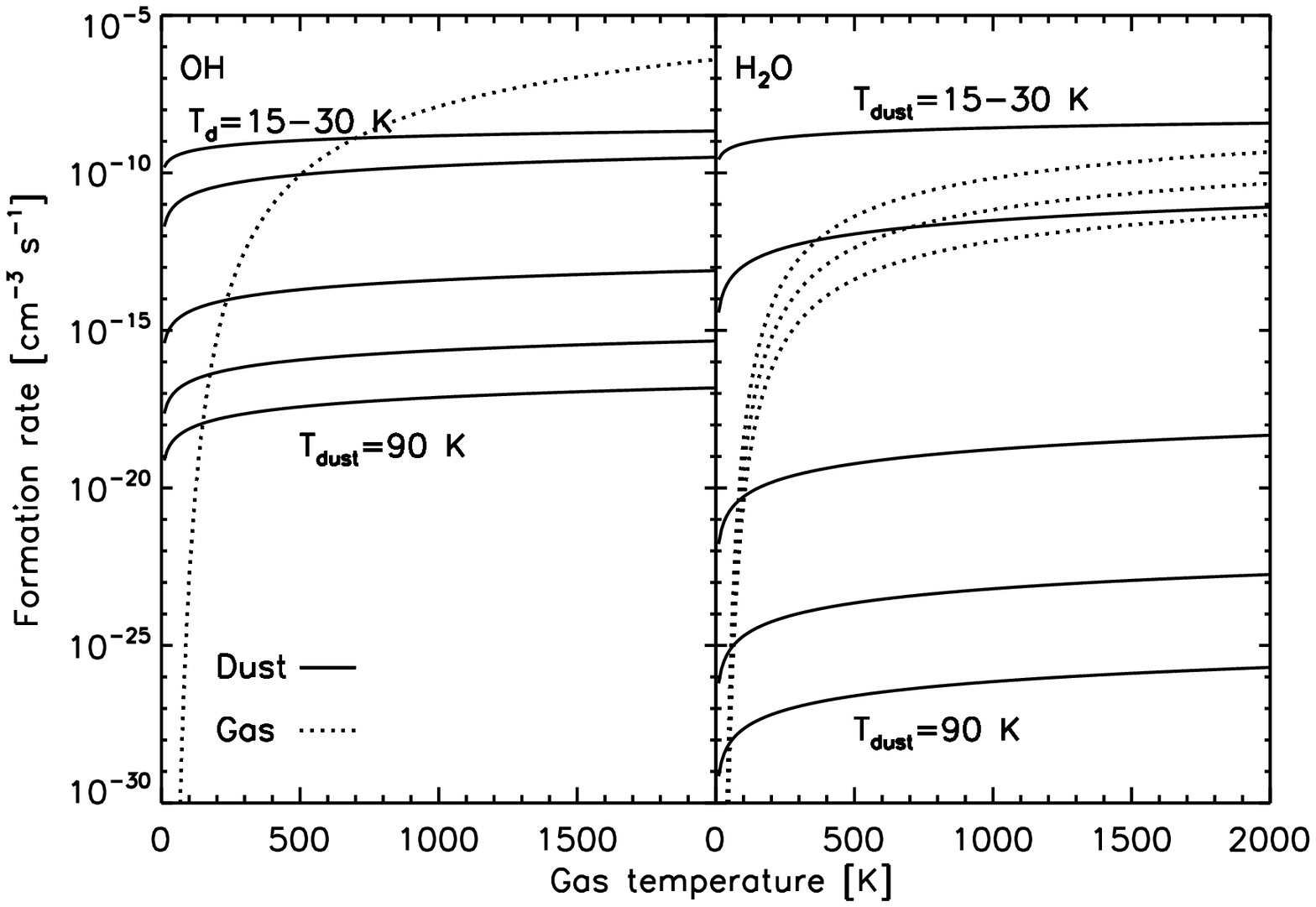}
  \includegraphics[width=8cm,clip]{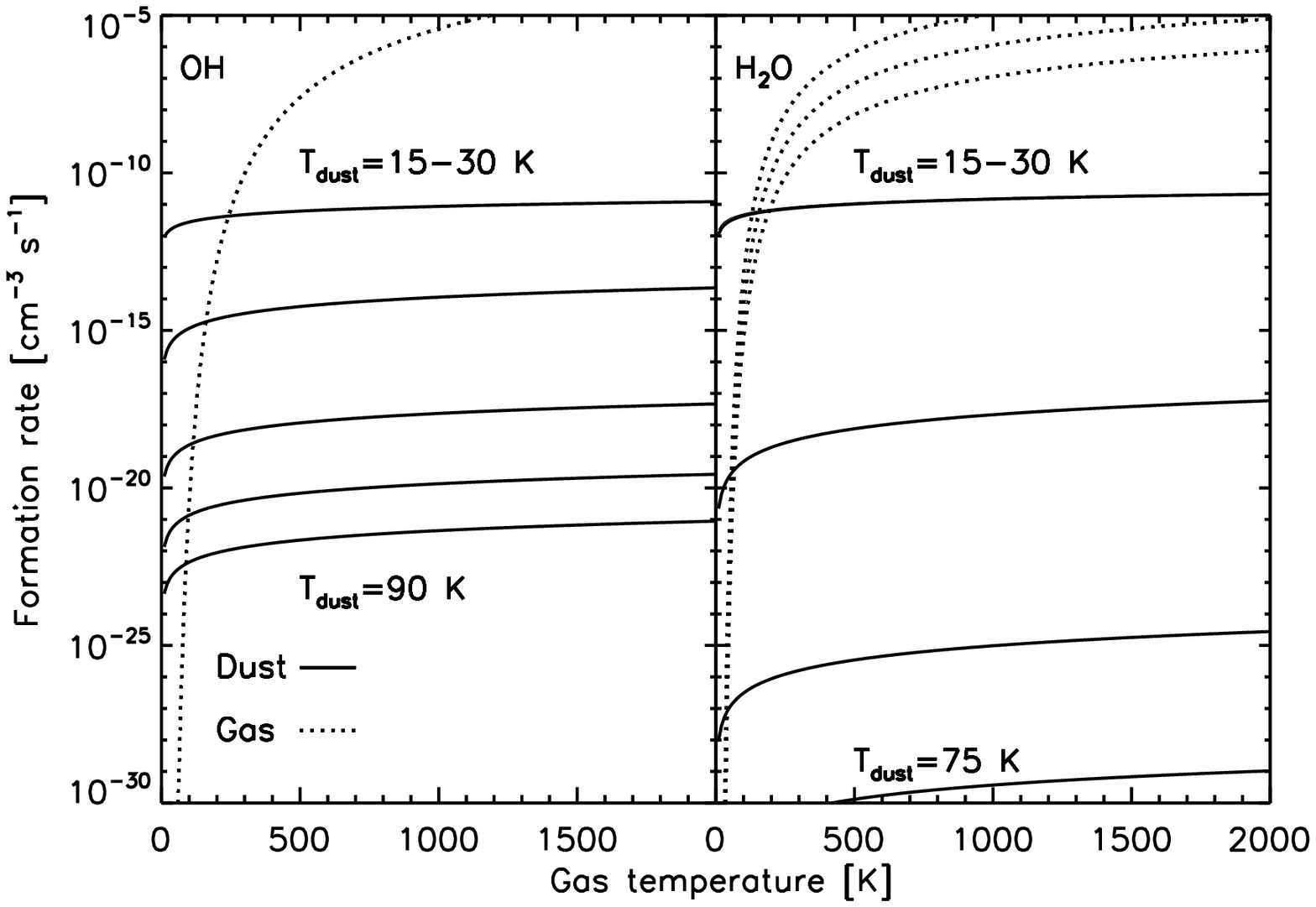}  \caption{Comparison of dust and gas phase reaction rates at
    $n=10^{5,5}$~cm$^{-3}$ for an environment with low (left, $x_{\rm
      H}=1$, $x_{\rm H_2}= 3\cdot 10^{-4}$) and high molecular
    abundances (right, $x_{\rm H}=10^{-2}$, $x_{\rm H_2}$=0.5) as a
    function of gas temperature for dust temperatures $T_{\rm
      dust}=15$, 30, 45, 60, 75 and 90~K. The OH abundances are varied
    between $x_{\rm OH}=10^{-8}$, $10^{-7}$, and $10^{-6}$ at low and
    $10^{-6}$, $10^{-5}$, and $10^{-4}$ at high molecular
    abundances. This resembles the unshielded and shielded environment
    of model 4 in \citet{Meijerink2005}.}\label{gas_phase_comparison}

\end{figure*}

\section{OH and H$_2$O formation: gas versus dust}

In order to estimate the importance of OH and H$_2$O formation on dust
grains, a comparison is made to the gas phase neutral-neutral
formation rates which are given by \citep[UMIST
  database:][]{LeTeuff2000,Woodall2007}:

\begin{eqnarray}
k_{\rm OH,gas} &=& 3.14\cdot 10^{-13} \left(\frac{T_{gas}}{300}\right)^{2.70} \exp(-\frac{3150}{T_{gas}})\ \ \rm cm^{3}~s^{-1} \\
k_{\rm H_2O,gas} &=& 2.05\cdot 10^{-12} \left(\frac{T_{gas}}{300}\right)^{1.52} \exp(-\frac{1736}{T_{gas}})\ \ \rm cm^{3}~s^{-1},
\end{eqnarray}

\noindent
while the conservative dust phase reaction rates (using the cross
sections from the MRN distribution, $n_{\rm dust}\sigma/n(\rm
H)=10^{-21}$~cm$^{-2}$), are given by:

\begin{eqnarray}
k_{\rm OH,dust} &=& 3.9 \cdot 10^{-17} \sqrt{\frac{T_{\rm gas}}{100}}\ \epsilon_{\rm OH}(T_{\rm dust})\ \ \rm cm^{3}~s^{-1}\\
k_{\rm H_2O,dust} &=& 3.9 \cdot 10^{-17} \sqrt{\frac{T_{\rm gas}}{100}}\ \epsilon_{\rm H_2O}(T_{\rm dust})\ \ \rm cm^{3}~s^{-1}.
\end{eqnarray}

\noindent
Depending on the environment (i.e., the ambient atomic and molecular
abundances, gas and dust temperature, etc.), OH and H$_2$O form
preferably on the dust grains or in the gas phase:

\begin{itemize}
\item Although the efficiency for formation of OH and H$_2$O on dust
  is strongly dependent on the dust temperature $T_{\rm dust}$, the
  variations with gas temperature $T_{\rm gas}$ at a fixed dust
  temperature $T_{\rm dust}$ are very modest.
\item The gas phase formation rates have large activation barriers of
  $T_{\rm gas}=3150$ and 1736~K for the formation of OH and H$_2$O,
  respectively. These channels are essentially cut off below
  temperatures $T_{\rm gas} \lesssim 200$~K.
\item Formation on dust depends on abundances of O and H, and in the
  gas phase on H$_2$, O, and OH.
\end{itemize}
For illustration purposes, a comparison between the gas and dust phase
formation rates are shown for two different regimes in
Fig. \ref{gas_phase_comparison}, a mostly mostly atomic (left) and
molecular (right) environment. We do not show the ion-molecule
reaction rates, even though they are considered in our chemical
network. These processes can be more efficient than the neutral
reaction chain at low gas temperatures, which is a domain that we do
not consider in this study. The abundunces resemble the unshielded and
shielded region of model 4 in \citet{Meijerink2005} with density
$n=10^{5.5}$~cm$^{-3}$: The atomic environment has abundances $x_{\rm
  H}=1$, $x_{\rm H_2}=3\cdot 10^{-4}$, $x_{\rm O}=3\cdot 10^{-4}$,
$x_{\rm OH}=10^{-8}$, $10^{-7}$, and $10^{-6}$ (left), and the
molecular environment has abundances $x_{\rm H}=10^{-2}$, $x_{\rm
  H_2}=0.5$, $x_{\rm O}=2\cdot 10^{-4}$, and $x_{\rm OH}=10^{-6}$,
$10^{-5}$, and $10^{-4}$ (right). We adopt a range of OH abundances,
to show the effect on the gas phase rates. We should note here that
the OSU astrochemistry
database\footnote{http://www.physics.ohio-state.edu/~eric/research.html}
(maintained by Eric Herbst) claims a smaller rate at high temperatures
$T > 500$ for the H$_2$ + OH reaction, and is given by $k_{\rm
  H_2O,gas}=8.4\times 10^{-13}\ \exp(-1040 / T_{\rm gas})$. This would
make the dust surface reaction even more important.

The efficiencies $\epsilon_{\rm OH}$ and $\epsilon_{\rm H_2O}$ are
reasonably high, $\sim 0.3$ and $\sim0.6$, respectively, at dust
temperatures $T_{\rm dust} < 40$~K. As a result, the H$_2$O dust
formation rate is always dominating the gas phase formation in the
atomic environment, while the OH dust formation is more important at
temperatures $T_{\rm gas} < 800$~K. Even though OH and H$_2$O
formation efficiencies drop very fast for higher dust temperatures,
molecule formation with dust as catalyst will always become dominant
below a certain gas temperature. The break even point ranges from
$T_{\rm gas}\sim 100 - 400$~K depending on the dust temperature
$T_{\rm dust}=45-90$~K. We should keep in mind though that for $T_{\rm
  dust}> 35-40$~K, the formation rates are overestimated in our model.

This picture is different in a molecular environment. Even when the OH
and H$_2$O formation efficiencies are high ($T_{\rm dust} < 40$~K, gas
phase formation is more efficient at gas temperatures $T_{\rm gas} >
100 - 200$~K. Where in the atomic environment, the OH and H$_2$O gas
phase formation is suppressed by the very low abundances of H$_2$, the
dust phase formation rate is suppressed due to the low accretion rate
of atomic hydrogen, which is depleted by two orders of magnitude. In
the molecular environment, the formation of water in either dust or
gas phase is inefficient through neutral-neutral reactions, and
ion-molecule reactions will also contribute significantly or even
dominate, through the chain $\rm O^+ + H_2 \rightarrow OH^+ +H$, $\rm
OH^+ + H_2 \rightarrow H_2O^+ +H$, $\rm H_2O^+ + H_2 \rightarrow
H_3O^+ +H$, followed by recombination $\rm H_3O^+ + e^- \rightarrow
H_2O + H$.

\section{Application to X-ray environments}

\begin{figure*}
  \centering
  \includegraphics[width=8cm,clip]{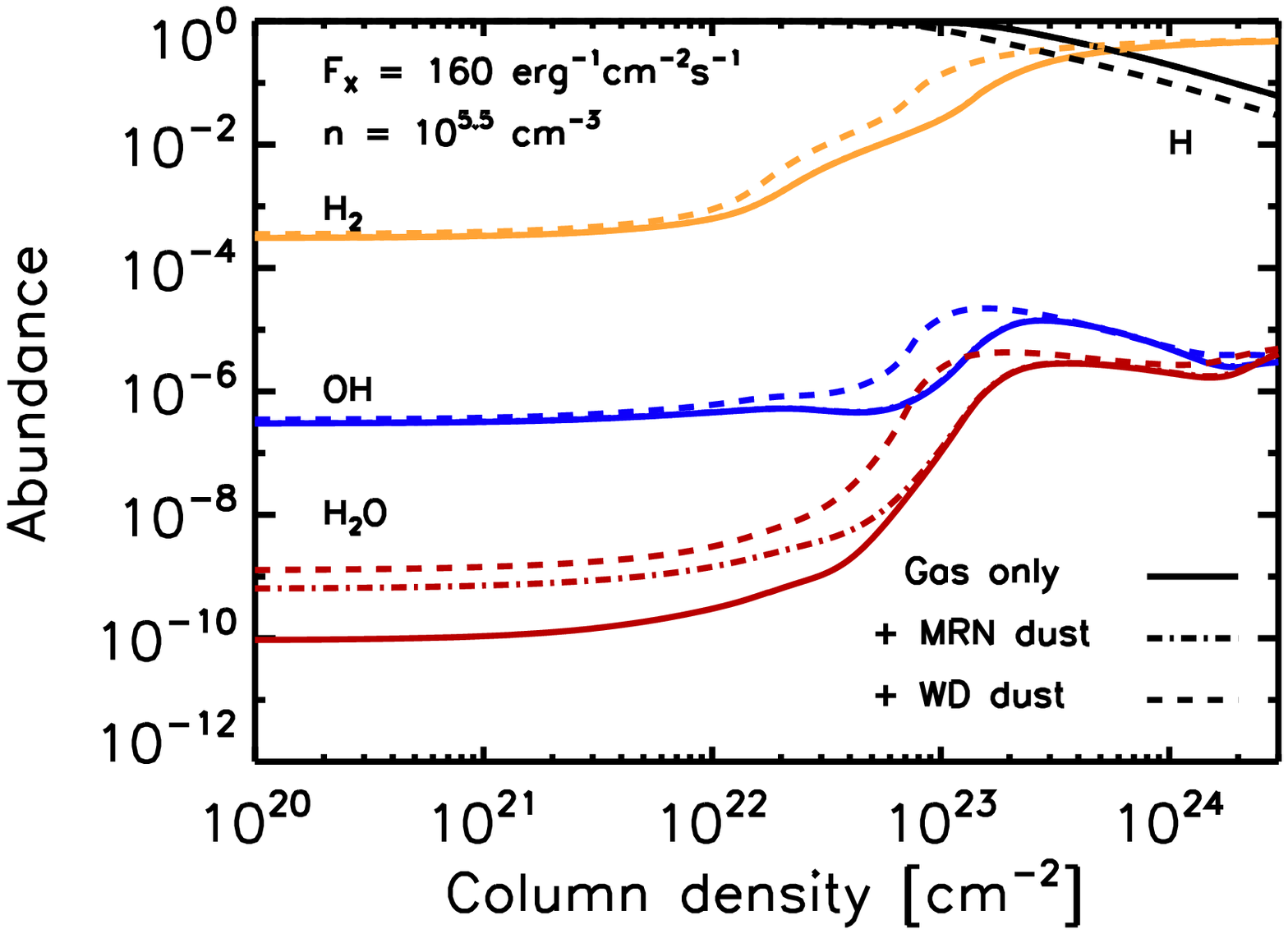}
  \includegraphics[width=8cm,clip]{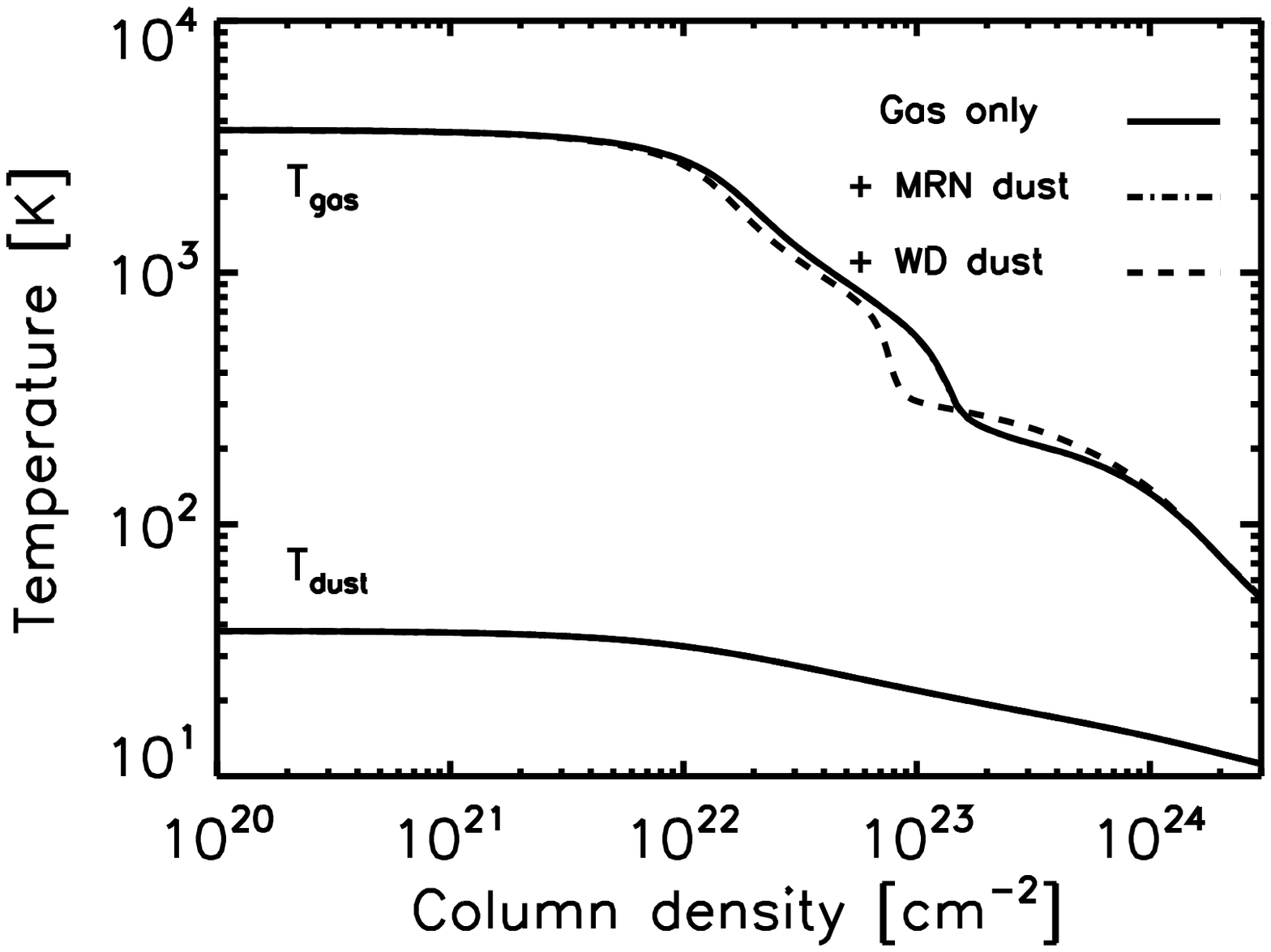}
  \caption{X-ray irradiated cloud with density $n=10^{5.5}$~cm$^{-3}$
    and radiation field $F_{\rm X}=160$~erg~cm$^{-2}$~s$^{-1}$. ({\it
      left}) The abundances are showns for H, H$_2$, OH, and H$_2$O in
    three different cases: formation of OH and H$_2$O (i) in gas phase
    only (ii) in gas phase and on dust with an MRN size distribution,
    and (iii) in gas phase and on dust with a Weingartner \& Draine
    size distribution. ({\it right}) Gas and dust temperatures for the
    same models. Note that the temperature profiles for the gas only
    models and those where OH and H$_2$O formation on dust with an MRN
    distribution are included are the same.}
  \label{XDRapplication}
\end{figure*}

The inner regions of active galaxies from, e.g., Mrk 231, which are
highly exposed to X-rays, are likely to exhibit enhanced formation of
H$_2$O on dust grains: (i) Dust temperatures are expected to be
moderate, $T_{\rm dust}\sim 20-40$~K, as the heating of dust is less
efficient than by UV; (ii) The dynamical timescales are not very long
and shocks are expected to be present, and therefore dust particles
are not expected to be or stay covered with ice; (iii) Shocks may even
be able to break up grains into smaller particles, therefore enlarging
the effective surface of dust, where water is expected to form.

In Fig. \ref{XDRapplication}, the chemical abundances of H, H$_2$, OH
and H$_2$O and temperatures of an XDR model is shown with density
$n=10^{5.5}$~cm$^{-3}$, $F_X=160$~erg~cm$^{-2}$~s$^{-1}$, and Solar
metallicity, $Z = Z_\odot$. Three different cases for the formation of
OH and H$_2$O are considered:

\begin{itemize}
\item[1.] Gas phase only: The OH abundance is $x_{\rm OH}\sim
  3\times10^{-7}$ at low column densities, rising to $\sim 10^{-5}$
  just before the H/H$_2$ transition, leveling of to $10^{-6}$. The
  H$_2$O abundance is $x_{\rm H_2O}\sim 10^{-10}$ at column densities
  $N_{\rm} < 10^{22}$~cm$^{-2}$. As OH, H$_2$O becomes more and more
  abundant toward the H/H$_2$ transition, and reaches an abundance
  $x_{\rm H_2O}\sim 10^{-6}$ beyond $N_{\rm H}\sim 10^{23}$~cm$^{-2}$.
\item[2.] Gas and dust with an MRN distribution: Nothing changes for
  OH. As shown in Fig. \ref{gas_phase_comparison}, the OH gas phase
  formation is more important at gas temperatures $T > 800$~K, at all
  dust temperatures. The abundance of H$_2$O, however, is higher by a
  factor 8, at low column densities $N_{\rm H} < 10^{22}$~cm$^{-2}$,
  and thus the integrated column of warm water ($T \gtrsim 200$~K). At
  low temperatures (i.e., at high column densities, $N_{\rm H} \gtrsim
  10^{23}$~cm$^{-2}$), nothing changes for OH and H$_2$O, as the
  formation route is dominated by the ion-molecule reactions.
\item[3.] Gas and dust with a Weingartner \& Draine distribution: The
  H$_2$ formation rate is now higher as well due to the higher surface
  area of the dust grains. As a result a small increase is seen in the
  H$_2$ abundance at low column densities.  At the same time, the
  H/H$_2$ transition occurs at lower column densities, $N_{\rm H} \sim
  2\times 10^{23}$~cm$^{-2}$, compared to $N_{\rm H} \sim 3\times
  10^{23}$~cm$^{-2}$ compared to the gas phase model. As a result, the
  OH peaks also at a lower column density, and reaches a slightly
  higher peak abundance. {\it The OH enhancement is due to a higher
    gas phase formation rate, as a result of the higher H$_2$
    abundance, not because of the higher formation rate on dust.}  The
  water abundance in the high temperature range is enhanced by more
  than an order of magnitude. As the H/H$_2$ transition occurs at
  smaller column densities, the rise of the water abundance also
  occurs at these smaller column densities.
\end{itemize}

To summarize, we find that the warm H$_2$O colunmn density is enhanced
by apprimately an order of magnitude over a column $N_{\rm H}\sim
5\cdot 10^{22}$~cm$^{-2}$, when including H$_2$O formation on dust. At larger
column densities, where temperatures decrease and the gas becomes
molecular, the formation on dust grains does not impact the abundance
of H$_2$O. The OH abundance is only indirectly affected by the higher
H$_2$ formation rate on dust when using the \citet{Weingartner2001}
distribution with the higher dust cross section.

\section{Discussion \& Conclusions}
The role and importance of OH and H$_2$O formation on dust is
investigated.  We use both Monte Carlo as well as rate equation
simulations to determine the efficiency at which oxygen is converted
into OH and H$_2$O on dust grains. The rate equation method is very
fast, and is used to calculate a grid of different environment, low
and high density ($n=10^3$ and $10^{5.5}$~cm$^{-3}$), different
molecular fractions, radiation fields, and temperatures (see
Figs. \ref{low_dens_eff} and \ref{high_dens_eff}).  The Monte Carlo
simulations are used as a consistency check, and as these simulations
are fairly slow, efficiencies are only calculated for a limited number
of dust temperatures. At dust temperatures $T_{\rm dust} < 35-40$~K,
the Monte Carlo and rate equations method yield the same results,
within the uncertainties.  At higher dust temperatures, the
efficiencies obtained by the two methods start to deviate: the rate
equation method systematically obtains larger efficiences than the MC
method. This is where the process of OH and H$_2$O formation on dust
enters the stochastic regime. We fitted analytical expression to the
derived results from the rate equations, such that they are easy to
implement in any chemistry code. We find that:

\begin{itemize}
\item OH formation on grains and release to gas phase is dominated by
 $\rm O + H \rightarrow OH$ ($T_{\rm dust} < 35-40$~K, and by $\rm O_3 +
 H \rightarrow OH + O_2$ (for higher dust temperatures). The high
 dust temperatures results in decreasing hydrogen surface densities,
 because of the increasing evaporation rates.
\item H$_2$O formation on grains and release to gas phase is dominated
 by $\rm OH + H \rightarrow H_2O$.
\item The fraction of OH and H$_2$O that is released in the gas upon
  formation is of the order of 36$\%$ and 9$\%$ respectively. However,
  in strong UV radiation fields, water on the dust can be
  photo-dissociated, and then reform to water, which then gives
  another contribution to the gas phase. This loop of
  forming-dissociating-forming.. increases the effective fraction of
  H$_2$O released in the gas by a factor 5.
\item A gas and dust phase formation rates comparison (see
Fig. \ref{gas_phase_comparison}) shows that formation on dust grains
is important for dust temperatures lower than 40~K.
\end{itemize}

We implemented the analytical expressions for the formation rates in
the \citet{Meijerink2005} XDR code, and calculated a model with
parameters representative for the inner regions of active galaxies. We
find that the additional processes enhances the integrated column
density of warm water. This increased column helps to explain
unusually strong water lines observed for Mrk 231, that has an
accreting supermassive black hole with X-ray luminosity $L_X \sim
10^{44}$~erg/s. This ULIRG shows for example a very high H$_2$O
$2_{02}-1_{11}$ / CO $J=8-7$ ratio. These ratios are not observed in
typical starforming environments that are mainly exposed to UV only
(PDR) environments, such as the one observed in the Orion Bar
\citep{Habart2010} or M82 \citep{Panuzzo2010}, where this ratio is
typically an order of magnitude or more smaller than in Mrk 231, and
as such these very bright H$_2$O lines might be typical for regions
that are exposed to high X-ray fluxes, close to the accreting black
hole in active galaxies.

\begin{acknowledgements}
We thank the anonymous referee for a careful reading of the manuscript
and giving a constructive report, that significantly improved the
paper.
\end{acknowledgements}

\bibliographystyle{aa}

\onecolumn

\begin{appendix}

\section{OH and H$_2$O formation efficiencies}\label{deriv_eff}

The formation efficiency with which gas phase OH and H$_2$O are formed
can be written as a balance between the desorption fraction from the
grain upon formation, and the accretion rate of oxygen atoms onto the
grain. On grains, two formation processes play a major role for OH in
different dust temperature regimes, while for H$_2$O, there is only
one major formation route:

\begin{eqnarray}
\rm O + H & \rightarrow & \rm OH  \ \ \ \ \ \ \ \ \ \ \ \ \ \ \ \ \ \ \ \ \ T_{\rm{dust}} < 40 K \label{OHformation_routes1} \\
\rm O_3 + H & \rightarrow & \rm OH + O_2  \ \ \ \ \ \ \ \ \ \ \ \ T_{\rm{dust}} > 40 K \label{OHformation_routes2} \\ 
\rm OH + H & \rightarrow & \rm H_2O \label{H2Oformation_route}
\end{eqnarray}

\noindent Therefore, the efficiencies should be written as:

\begin{eqnarray}
\epsilon_{\rm OH} &=& \frac{f_{des_{\rm OH}} n({\rm O}) n(\rm H) \alpha_{\rm H}}{R_{acc}({\rm O})} +
\frac{f_{des_{\rm OH'}} n({\rm O_3}) n({\rm H}) \alpha_{\rm H}}{R_{acc}({\rm O})}\label{OHeff} \\
\epsilon_{\rm H_2O} &=& \frac{f_{des_{\rm H_2O}} n({\rm OH}) n(\rm H) \alpha_{\rm H}}{R_{acc}({\rm O})}  \label{H2Oeff}
\end{eqnarray}

\noindent
$n({\rm H})$, $n(\rm O)$, $n(\rm O_3)$, and $n(\rm OH)$ are the
surface density of H, O, O$_3$, and OH on the dust. These hydrogen and oxygen atoms are
accreted from the gas phase at a rate:

\begin{eqnarray} 
R_{acc}({\rm H})& =& \frac{n_{\rm HI}v_{\rm HI}}{N_S} {\rm\ monolayers}\ s^{-1}, \\
R_{acc}({\rm O})& =& \frac{n_{\rm OI}v_{\rm HI}}{N_S} {\rm\ monolayers}\ s^{-1},
\end{eqnarray}

\noindent where $n_{\rm HI}$ and $n_{\rm OI}$ are the densities of
atomic hydrogen and oxygen in the gas phase, $v_{\rm HI}$ and $v_{\rm
  OI}$ the mean velocities of these atoms:

\begin{eqnarray}
v_{\rm HI}&\sim& 1.45\cdot 10^5 \sqrt{\frac{T_{\rm gas}}{100}}\ {\rm cm}\ {\rm s}^{-1},\\
v_{\rm OI}&\sim & 3.6\cdot 10^4 \sqrt{\frac{T_{\rm gas}}{100}}\ {\rm cm}\ {\rm s}^{-1},
\end{eqnarray}

\noindent
and $N_S=\frac{1}{a^2_{pp}}=8\cdot 10^{14}$~cm$^{-2}$ the number
  of sites per cm$^2$. H atoms arrive on the dust surface in
  physisorbed sites, and then continue to a chemisorbed site or
  evaporate from the surface. As a result, the density on the surface
  can be written as

\begin{equation}
n({\rm H}) = \frac{R_{acc}({\rm H})}{\alpha_{pc} + R_{evap_{\rm H}} }
\end{equation}

\noindent
The mobility for hydrogen atoms to enter into a chemisorbed site is
mostly due to tunneling, which we can then approximate by
$\alpha_{pc}=10^{12} \exp(-2a_{pc}\sqrt{\frac{2m_{\rm
      H}E_A}{\hbar^2}})=6.7\times 10^{-3}$, where $a_{pc}$ and $E_A$
are the width and the height of the barrier between physisorbed and
chemisorbed sites. The adopted values are $a_{pc} = 1.5$~\AA\ and
$E_A=2870$~K. The hydrogen surface density can then be calculated as
follows

\begin{equation}
n({\rm H}) = \frac{1.45\cdot 10^{-10} n_{\rm HI} \sqrt{\frac{T_{\rm gas}}{100}}}{6.7\times 10^{-3} + 10^{12}\exp(\frac{-526}{T_{\rm dust}})}
\end{equation}

\noindent
The formation of OH and H$_2$O can be determined analytically from the
rate equations, when the most important processes are considered. The
rate equations of O, OH, and H$_2$O, as well as O$_2$ and O$_3$ are
considered here, since the efficiency of OH depends on the abundance
of O$_3$ as well. The abundance change as a function of time for these
species are given by by the following expressions:

\begin{eqnarray}
\frac{dn({\rm O})}{dt} &=& R_{acc}({\rm O}) - n({\rm O}) n({\rm H}) \alpha_{\rm H} + n({\rm OH}) R_{phot_{\rm OH}} -
2 n(O)^2 \alpha_{\rm O} - R_{evap_{\rm O}} n({\rm O}) \\
\frac{dn({\rm OH})}{dt} &=& (1 - f_{des_{\rm OH}})n({\rm O}) n({\rm H}) \alpha_{\rm H} + R_{phot_{\rm H_2O}} n({\rm H_2O})-  R_{phot_{\rm OH}} n({\rm OH}) - n({\rm OH}) n({\rm H}) \alpha_{\rm H} -  R_{evap_{\rm OH}} n({\rm OH}) \\
\frac{dn({\rm H_2O})}{dt} &=& (1 - f_{des_{\rm H_2O}})n({\rm OH}) n({\rm H}) \alpha_{\rm H}-  R_{phot_{\rm H_2O}} n({\rm H_2O}) - n({\rm H_2O}) n({\rm H}) \alpha_{\rm H} \\
\frac{dn({\rm O_2})}{dt} &=& (1 - f_{des_{\rm O_2}}) n({\rm O})^2 \alpha_{\rm O} + (1 - f'_{des_{\rm O_2}}) n({\rm OH}) n({\rm O}) \alpha_{\rm OH} - n({\rm O_2}) n({\rm H}) \alpha_{\rm H} - R_{phot_{\rm O_2}} n({\rm O_2}) - R_{evap_{\rm O_2}} n({\rm O_2}) \\
\frac{dn({\rm O_3})}{dt} &=&  (1 - f_{des_{\rm O_3}}) n({\rm O_2}) n({\rm O}) \alpha_{\rm O} - n({\rm O_3}) n({\rm H}) \alpha_{\rm H} - n({\rm O_3}) n({\rm OH}) \alpha_{\rm OH} - R_{phot_{O_3}} n({\rm O_3})
\end{eqnarray}

\noindent
When we consider the surface densities of the species at steady state,
the surface densities of the species become

\begin{eqnarray}
n({\rm O}) &=& \frac{R_{acc}({\rm O})}{n(H) \alpha_{\rm H} + R_{evap_{\rm O}} - R_{phot_{\rm OH}}\frac{1 - f_{des_{\rm OH}}}{f_{des_{\rm H_2O}} + \frac{R_{phot_{\rm OH}}}{n({\rm H})\alpha_{\rm H}}} + \sqrt{2 R_{acc}({\rm O}) \alpha_{\rm O}}} \\
n({\rm OH}) &=& \frac{(1 - f_{des_{\rm OH}})n({\rm O})}{f_{des_{\rm H_2O}} + \frac{R_{phot_{\rm OH}}}{n({\rm H})\alpha_{\rm H}} + \frac{R_{evap_{\rm OH}}}{n({\rm H})\alpha_{\rm H}}} \\
n({\rm H_2O}) &=& \frac{(1 - f_{des_{\rm H_2O}}) n({\rm OH}) n({\rm H}) \alpha_{\rm H}}{n({\rm H}) \alpha_{\rm H} + R_{phot_{\rm H_2O}}} \\
n({\rm O_2}) &=& \frac{(1-f_{des_{O_2}}) n({\rm O})^2 \alpha_{\rm O}  + (1 - f'_{des_{O_2}}) n({\rm OH}) n({\rm O}) \alpha_{\rm OH}}{{n({\rm H}) \alpha_{\rm H} + R_{phot_{\rm O_2}} + R_{evap_{\rm O_2}}}} \\
n({\rm O_3}) &=&  \frac{(1-f_{des_{O_3}}) n({\rm O_2}) n({\rm O}) \alpha_{\rm O}}{n({\rm H}) \alpha_{\rm H} + n(\rm OH) \alpha_{\rm OH} + R_{phot_{\rm O_3}}}
\end{eqnarray}

Filling in the equations for the various included reactions, the expressions become as follows:

\begin{eqnarray}
n({\rm O}) &=& \frac{R_{acc}({\rm O})}{10^{12} \exp(\frac{-350}{T_{\rm dust}}) n({\rm H}) + 10^{12} \exp(\frac{-1366}{T_{\rm dust}}) - \frac{3.90\cdot 10^{-10} G_0 \cdot 0.64}{0.15 + \frac{3.9\cdot 10^{-10} G_0}{n({\rm H}) \exp(\frac{-350}{T_{\rm dust}})}} + \sqrt{R_{acc}({\rm O}) 10^{12} \exp(\frac{-910}{T_{dust}})} } \nonumber \\ 
&=& \frac{10^{-12} R_{acc}({\rm O})}{n({\rm H})\exp(\frac{-350}{T_{\rm dust}}) + \exp(\frac{-1366}{T_{\rm dust}}) - \frac{2.5\cdot 10^{-22} G_0}{0.15 + 3.9\cdot 10^{-22} G_0 \exp(350 / T_{dust}) / n({\rm H})} + \sqrt{2\cdot 10^{-12} R_{acc}({\rm O}) \exp(\frac{-910}{T_{dust}})} } \\
n({\rm OH}) &=& \frac{0.64 n({\rm O})}{0.15 + \frac{3.90\cdot 10^{-10} G_0}{n({\rm H}) 10^{12} \exp(\frac{-350}{T_{dust}})} + \frac{10^{12} \exp(\frac{-1336}{T_{dust}})}{n({\rm H}) 10^{12} \exp(\frac{-350}{T_{dust}})}}  \nonumber \\
&=& \frac{0.64 n({\rm O}) n({\rm H})}{0.15 n({\rm H}) + 3.9\cdot 10^{-22} G_0 \exp(\frac{350}{T_{dust}}) + \exp(\frac{-986}{T_{dust}})} \\
n({\rm H_2O}) &=& \frac{0.85 n({\rm H}) n({\rm OH}) 10^{12} \exp(\frac{-350}{T_{dust}})}{n({\rm H}) 10^{12} \exp(\frac{-350}{T_{dust}}) + 8 \cdot 10^{-10} G_0} \nonumber \\
&=&  \frac{0.85 n({\rm H}) n({\rm OH}) \exp(\frac{-350}{T_{dust}})}{n({\rm H}) \exp(\frac{-350}{T_{dust}}) + 8 \cdot 10^{-22} G_0} \\
n({\rm O_2}) &=& \frac{0.975 n({\rm OH}) n({\rm O}) 10^{12} \exp(\frac{-890}{T_{dust}}) + 0.64 n({\rm O_2}) 10^{12} \exp(\frac{-910}{T_{dust}})}{ n({\rm H}) 10^{12} \exp(\frac{-350}{T_{dust}}) + 7.90\cdot 10^{-10} G_0 + 10^{12} \exp(\frac{-1416}{T_{dust}}) } \nonumber\\
&=& \frac{0.975 n({\rm OH}) n({\rm O}) \exp(\frac{-890}{T_{dust}}) + 0.64 n({\rm O_2}) \exp(\frac{-910}{T_{dust}})}{ n({\rm H}) \exp(\frac{-350}{T_{dust}}) + 7.90\cdot 10^{-22} G_0 + \exp(\frac{-1416}{T_{dust}}) } \\
n({\rm O_3}) &=& \frac{ 0.977 n({\rm O}) n({\rm O_2}) 10^{12} \exp(\frac{-910}{T_{dust}})}{n(\rm H) 10^{12} \exp(\frac{-350}{T_{dust}}) + 10^{12} \exp(\frac{-890}{T_{dust}}) n({\rm OH}) + 1.8\cdot 10^{-10} G_0} \nonumber\\
 &=& \frac{ 0.977 n({\rm O}) n({\rm O_2}) \exp(\frac{-910}{T_{dust}})}{n(\rm H) \exp(\frac{-350}{T_{dust}}) + \exp(\frac{-890}{T_{dust}}) n({\rm OH}) + 1.8\cdot 10^{-22} G_0}
\end{eqnarray}

\begin{table}
\caption{Adopted photodissociation rates in the approximation}   
\label{important_reactions}   
\centering                    
\begin{tabular}{c c }      
\hline\hline                
Reaction                        & Rate (s$^{-1}$) \\   
\hline                        
OH     + photon $\rightarrow$ O     + H  & 3.9(-10) \\
O$_2$  + photon $\rightarrow$ O     + O  & 7.9(-10) \\      
H$_2$O + photon $\rightarrow$ OH    + H  & 8.0(-10) \\
O$_3$  + photon $\rightarrow$ O$_2$ +O   & 1.9(-09) \\
\hline                                  
\end{tabular}
\label{photorates}
\end{table}

\end{appendix}

\end{document}